\documentclass[sigplan,screen]{acmart}

\usepackage{subcaption}%
\usepackage{filecontents}
\usepackage{nameref}
\usepackage{tikz}
\usepackage{amsmath}
\usepackage{balance}
\usepackage{titlesec}%
\makeatletter 
\newcommand\semilarge{\@setfontsize\semilarge{11}{12}}
\makeatother
\titleformat*{\subsection}{\semilarge\bfseries}

\usepackage{enumitem}
\usepackage{pbox}
\usepackage{multirow}
\usepackage{tabularx}
\usepackage{makecell}
\usepackage{mathtools}
\usepackage{caption}
\usepackage{float}
\usepackage{booktabs}
\usepackage[linesnumbered,ruled,vlined]{algorithm2e}
\usepackage{algorithm2e}%
\usepackage{xcolor}
\usepackage{nameref}
\usepackage{wrapfig}
\usepackage{xspace}
\usepackage{pifont}

\copyrightyear{2025}
\acmYear{2025}
\setcopyright{rightsretained}
\acmConference[ASPLOS '25] {Proceedings of the 30th ACM International Conference on Architectural Support for Programming Languages and Operating Systems, Volume 1}{March 30--April 3, 2025}{Rotterdam, Netherlands.}
\acmBooktitle{Proceedings of the 30th ACM International Conference on Architectural Support for Programming Languages and Operating Systems, Volume 1 (ASPLOS '25), March 30--April 3, 2025, Rotterdam, Netherlands}
\acmISBN{979-8-4007-0698-1/25/03}
\acmDOI{10.1145/3669940.3707244}
\acmISBN{979-8-4007-0698-1/25/03}



\begin{document}
\newcommand{\res}{{Phoenix}\xspace}
\newcommand{\cats}{{CAP}\xspace}
\newcommand{\tags}{{Resilience Tags}\xspace}
\newcommand{\rewrite}[1]{{{\textcolor{blue}{[Rewrite: #1]}}}}
\newcommand{\kubepri}{{Bestfit\_Pri}\xspace}
\newcommand{\kubefair}{{Bestfit\_Fair}\xspace}

\newcommand{\parab}[1]{\vspace{0.05in}\noindent\textbf{#1}}
\newcommand{\parait}[1]{\vspace{0.05in}\noindent\textit{#1}}

\newcommand{\fix}[1]{\textcolor{red}{#1}}
\newcommand{\kapil}[1]{\textcolor{black}{#1}}
\newcommand{\saj}[1]{\textcolor{black}{#1}}

\newcommand{\cmark}{\ding{51}}
\newcommand{\xmark}{\ding{55}}

\title[Cooperative Graceful Degradation In Containerized Clouds]{Cooperative Graceful Degradation In \\Containerized Clouds}

\author{Kapil Agrawal}
\affiliation{%
  \institution{University of California, Irvine}
  \country{}
}
\email{kapila1@uci.edu}

\author{Sangeetha Abdu Jyothi}
\affiliation{%
  \institution{University of California, Irvine \& VMware Research}
  \country{}
}
\email{sangeetha.aj@uci.edu}

\begin{abstract}
Cloud resilience is crucial for cloud operators and the myriad of applications that rely on the cloud. Today, we lack a mechanism that enables cloud operators to perform graceful degradation of applications while satisfying the application's availability requirements. In this paper, we put forward a vision for automated cloud resilience management with cooperative graceful degradation between applications and cloud operators. First, we investigate techniques for graceful degradation and identify an opportunity for cooperative graceful degradation in public clouds. Second, leveraging criticality tags on containers, we propose diagonal scaling---turning off non-critical containers during capacity crunch scenarios---to maximize the availability of critical services. Third, we design Phoenix, an automated cloud resilience management system that maximizes critical service availability of applications while also considering operator objectives, thereby improving the overall resilience of the infrastructure during failures. We experimentally show that the Phoenix controller running atop Kubernetes can improve critical service availability by up to $2\times$ during large-scale failures. Phoenix can handle failures in a cluster of 100,000 nodes within 10 seconds. We also develop AdaptLab, an open-source resilience benchmarking framework that can emulate realistic cloud environments with real-world application dependency graphs.

\end{abstract}
\maketitle

\section{Introduction}
\label{sec:intro}
Public and private clouds host a myriad of applications from diverse domains. The resilience of the cloud infrastructure is crucial for maintaining the business continuity of these applications. However, as cloud infrastructure expands, the number of infrastructure incidents has also been rising significantly~\cite{lee2021shard, meta-cap-crunch-blog, veeraraghavan2018maelstrom}, with nearly 40\% of production incidents caused by infrastructure failures~\cite{ghosh2022fight}. The recovery time of such infrastructure incidents is typically long, as it involves work by on-site personnel~\cite{veeraraghavan2018maelstrom}, leading to massive user-visible outages and revenue loss~\cite{meta-cap-crunch-blog,aws-revenue-outage, dc-heatwave, aws-outage}. 

Cloud operators and applications employ a variety of solutions to improve availability during failures. The solutions on the \textit{operator front} can be broadly classified into three categories based on when they are employed---proactive, mitigation, and recovery. Proactively, large-scale cloud providers conduct risk assessment~\cite{alipourfard2019risk, xia2021social, failure-modes} and disaster-driven capacity planning~\cite{keeton2004designing,eriksen2023global, zhu2021network,newell2021ras}, typically adding sufficient redundancy across all components of the infrastructure~\cite{patterson1988case, lee2021shard, tia-942}. Cloud providers also perform disaster audits proactively, using stress tests to identify vulnerabilities in the infrastructure~\cite{dirt, netflix-chaos, awsgameday}. During failures, cloud providers employ a range of mitigation solutions to minimize the impact~\cite{failover, lee2021shard, levy2020predictive, veeraraghavan2018maelstrom,lyu2023hyrax, kumbhare2021prediction, li2020thunderbolt, wu2012netpilot, piga2024expanding, wu2016dynamo, krishnaswamy2022decentralized}. For example, cloud operators maintain rule-based runbooks with tasks that must be performed during disaster scenarios (migration of data, restarting containers, etc.)~\cite{veeraraghavan2018maelstrom}. Post-disaster recovery include damage assessment~\cite{barroso2018datacenter}, on-site repairs~\cite{veeraraghavan2018maelstrom}, and restoration~\cite{minh2014fly, minh2016site,geng2015lessons}.

On the \textit{application front}, the key goal is graceful degradation during failures, also referred to as self-adaptation~\cite{ klein2014brownout, lattice, fox1999harvest, zaharia2012resilient, meza2023defcon, dean2008mapreduce, app-graceful1, app-graceful2, web-graceful1, web-graceful2, zhou2018overload}. Several application-level resilience products~\cite{springboot, resilience4j, gobackoff, golimiter, hystrix, istio, sentinel, grpc-fault-tolerance, wire, jepsen} that applications can readily incorporate to contain failures provide out-of-box capabilities, such as circuit-breakers~\cite{hystrix} and rate-limiters~\cite{golimiter}. Furthermore, open-source chaos testing tools~\cite{chaos-monkey, litmus-chaos, gremlin} can proactively test the efficacy of resilience patterns to detect any undesirable behaviors.

Today, cloud resilience management mainly involves independently operating solutions at the operator and application levels, particularly in public clouds. Operator solutions typically treat applications as blackboxes~\cite{levy2020predictive, kumbhare2021prediction} and rely only on infrastructure-level signals to mitigate the impact of failures. For example, a recent solution~\cite{kumbhare2021prediction} identifies VMs that can be potentially throttled by inferring their criticality through analysis of access patterns. While choosing mitigation actions, another system, Narya~\cite{levy2020predictive}, assumes that a single long VM outage is preferable to applications over multiple short ones. These black-box solutions that are agnostic of application requirements could inadvertently terminate critical components of the application and affect its availability significantly. Thus, the current practice of employing independent resilience solutions at infrastructure and application levels hurts overall availability.

We ask an alternative question: Is cooperative graceful degradation feasible in public clouds while maintaining applications' black-box nature? If applications could share their resilience requirements without divulging business logic, operators could incorporate these requirements into infrastructure-level resilience solutions. Recently, Meta~\cite{meza2023defcon} demonstrated the benefits of a cooperative degradation approach in their \textit{private} cloud, leveraging visibility and control over both applications and infrastructure. In Defcon~\cite{meza2023defcon}, application developers incorporate knobs to expose degradable features controlled by the infrastructure, allowing non-critical features to be turned off during capacity crunch scenarios. However, this approach is not feasible in public clouds, as it requires application modifications.

We argue that cooperative graceful degradation is indeed feasible in public clouds without compromising the black-box nature of applications. We examine the design space for realizing graceful degradation (\S~\ref{sec:ds}) and find that prior solutions have explored the two extremes, i.e., requiring fine-grained application changes~\cite{meza2023defcon} or treating the application as a complete black-box~\cite{levy2020predictive,kumbhare2021prediction}. We make the case that exposing the resilience requirements at the \textit{container level} gives operators better visibility without revealing the application's business logic. Thus, the widely adopted container paradigm serves as an ideal abstraction for implementing cooperative graceful degradation. Furthermore, in microservice-based applications, where functionality is decomposed into independently developed and deployed containerized microservices, container-level degradation is a natural fit. 

In this paper, we present Phoenix, a cooperative graceful degradation framework for containerized clouds hosting microservice-based applications. Phoenix's key goal is to satisfy application-level resilience requirements maximally while also considering operator objectives and resource availability in capacity-constrained cloud environments following failures. On the application front, Phoenix relies on a simple and expressive abstraction to convey the application's resilience requirements---\textit{Criticality Tags} on containers. On the operator front, the Phoenix automated resilience management system converts application-level criticality tags and operator-level objectives to actionable capacity reallocation decisions for cluster schedulers~\cite{tirmazi2020borg, tang2020twine, kubernetes}.

At its core, Phoenix comprises a criticality-aware planner and scheduler. During a failure event, the planner takes as input the container information of applications, along with their criticality tags, to generate a prioritized list of microservices to activate within the available capacity. This process has two steps. First, the planner creates an ordered list of containers at the per-application level based on criticality tags. Second, it ranks containers across applications based on cloud operator objectives, such as fairness or revenue targets. Based on the planner’s globally ordered list, the scheduler then generates a sequence of actions (including scheduling, migrating, or shutting down) for running microservices at remaining healthy servers.

Cooperative graceful degradation with Phoenix offers several benefits. First, a fast response is critical during disaster events. A centralized resilience mechanism with access to application-level criticality information can speed up the response time and improve availability. Second, by specifying their acceptable degraded states using various levels of criticality to the cloud---beyond the commonly-used notion of the entire application being "on" or "off"---applications can transform their resilience objectives from a single scalar value to a range of potential values. For example, the Recovery Time Objective (RTO), a commonly used resilience objective defined as the maximum acceptable time an application can be unavailable, may be expanded to include intermediate RTO requirements for degraded states. Critical sub-services of an application can have stringent RTO bounds, while non-critical sub-services may allow more flexibility. Finally, at the infrastructure level, several large-scale failures within a data center, which would otherwise require failing over to a backup data center, can be handled in place. 

We deploy Phoenix on a Kubernetes cluster running \textit{Overleaf}~\cite{overleaf}, a real-world microservice-based document editing application, and the Hotel Reservation application from DeathStarBench~\cite{gan2019open} to demonstrate the feasibility and benefits of cooperative degradation. \footnote{Note that Phoenix is agnostic to the underlying cluster scheduler and can support other schedulers~\cite{hindman2011mesos, tirmazi2020borg,tang2020twine,vavilapalli2013apache}.} We also develop AdaptLab, a resilience benchmarking platform, to emulate realistic large cloud environments of sizes up to 100,000 nodes running real-world microservice application dependency graphs obtained using Alibaba traces. Benchmarking results show that Phoenix's cooperative graceful degradation can maximize critical service availability across applications while also satisfying operator objectives, outperforming non-cooperative baselines. Our AdaptLab simulations show that the planning time of Phoenix is under 10 seconds for a 100,000-node cluster. On a real-world 200 CPU Kubernetes cluster, we observe that the end-to-end time taken by Phoenix to provide full recovery for all applications---including deletions, restarts, etc.---is under 4 minutes. 

While Phoenix takes the first step towards practical cooperative degradation in large-scale public clouds, the focus in this paper is on stateless workloads, which account for over 60\% of resource utilization in large data centers~\cite{lee2021shard}. Extending cooperative degradation to stateful workloads requires substantial research and is left as future work.

In summary, we make the following contributions:
\begin{itemize}[leftmargin=*,nolistsep]
    \item We make a case for cooperative graceful degradation in public cloud environments. 
    \item We develop Phoenix, an automated cloud resilience management system that employs resilience-aware planning and scheduling to maximize both application-level resilience and operator objectives during capacity-constrained failure scenarios.
    \item We develop AdaptLab, a resilience benchmarking platform that can emulate disasters of varying failure rates in realistic cloud environments with real-world microservice workloads. 
    \item Phoenix controller running two microservice applications, Overleaf~\cite{overleaf} and HotelReservation from DeathStarBench \cite{gan2019open}, in a 200 CPU Kubernetes cluster improves critical service availability by up to $2\times$ under large-scale failures and provides full recovery in under 4 minutes.
    \item Phoenix can generate new plans for clusters with up to 100,000 nodes in under 10 seconds.

\end{itemize}

\section{Background and Related Work}
\label{subsec:motivate}

Ensuring cloud resilience under extreme events whose time of occurrence, location, duration, and strength may be difficult to predict is a challenge for cloud operators. 

\parab{Threats:} There exist several planned and unplanned threats that affect the availability of cloud resources. This includes natural threats such as severe weather events~\cite{severe-weather1,severe-weather2}, including heatwaves~\cite{dc-heatwave}, hurricanes~\cite{veeraraghavan2018maelstrom}, floods~\cite{dc-failure-modes}, fires~\cite{failure-modes}, and, in the extreme case, solar storms~\cite{solarstorm}. Human errors~\cite{aws-outage} and faulty operation of automated systems~\cite{catchpoint, ghosh2022fight} can also result in large-scale outages in the cloud. In addition to extraneous threats, during its normal operation, the cloud may experience unexpected load spikes, leading to power outages~\cite{kumbhare2021prediction, li2020thunderbolt, wu2016dynamo}, hardware failures~\cite{gunawi2014bugs, gunawi2016does, lyu2023hyrax, levy2020predictive, wu2012netpilot}. On longer timescales, the process of adding new equipment to the cloud may not keep up with the rate of increase in load~\cite{xia2021social}. In short, several factors pose a risk to the reliability and availability of cloud services. We refer to a large-scale failure due to any of the above causes as a ``disaster''.

\begin{figure}[t!]
	\centering
	\includegraphics[width=0.9\linewidth]{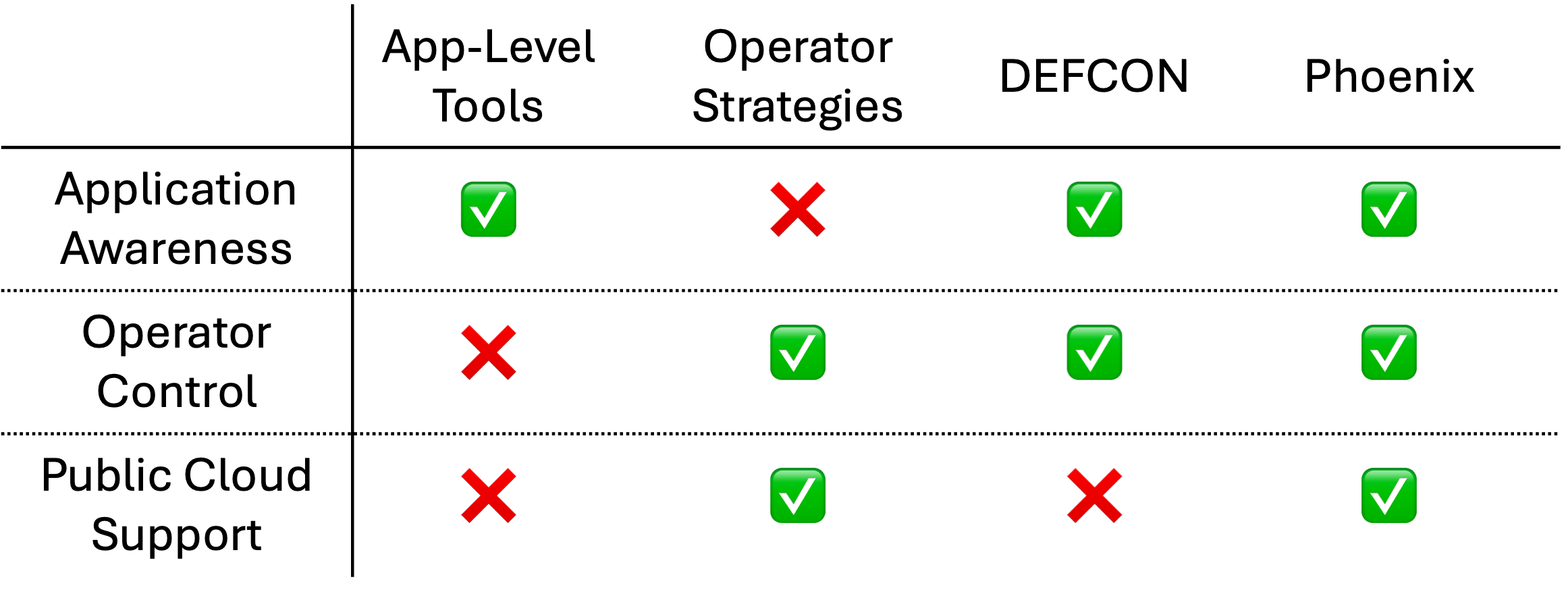}
	\caption{\small Table comparing features of app-level tools~\cite{xu2019brownout, klein2014brownout,resilience4j, hystrix, golimiter, gobackoff, istio, consul,web-graceful1, web-graceful2, mail-graceful1, storage-graceful1, storage-graceful2, storage-graceful3, search-graceful1,load-shedding1, load-shedding2, load-shedding3,beyer2016site, cho2020overload, zhou2018overload}, operator strategies~\cite{levy2020predictive,kumbhare2021prediction}, DEFCON~\cite{meza2023defcon} and Phoenix.} 
	\label{fig:compare}
 \vspace{-12pt}
\end{figure}

\parab{Cloud Resilience Solutions:} 
Infrastructure-level resilience solutions can be broadly classified into pro-active (before disaster), mitigation (during disaster), and recovery (post-disaster). Pre-disaster resilience solutions include risk assessment~\cite{alipourfard2019risk, xia2021social} to assess failure modes~\cite{failure-modes}, capacity planning ~\cite{ahuja2021capacity, newell2021ras, eriksen2023global} by adding sufficient redundancy on components such as power distribution units, cooling units, servers, and network~\cite{tia-942, barroso2018datacenter, denis2023ebb}. Several large cloud providers have in-house disaster readiness teams~\cite{dirt, netflix-chaos} which conduct regular disaster audits~\cite{awsgameday} by stress testing all layers of the cloud stack. During disaster, mitigation actions include failover~\cite{lee2021shard}, enabling new capacity~\cite{alipourfard2019risk}, soft reboots~\cite{levy2020predictive}, fail in-place~\cite{lyu2023hyrax}, load-shaping~\cite{li2020thunderbolt, kumbhare2021prediction}, etc. For large-scale failures such as hurricanes, data center draining~\cite{veeraraghavan2018maelstrom, dc-shutdown-dropbox} have been proposed. Finally, post-disaster solutions include on-site repairs~\cite{veeraraghavan2018maelstrom} and recovery~\cite{minh2014fly, minh2016site}, to name a few.

\parab{Graceful Degradation:} Our focus is on a class of mitigation solutions known as graceful degradation~\cite{fox1999harvest}, where systems are designed to continue functioning with reduced performance or limited features during capacity-crunch scenarios. This is also referred to as self-adaptation~\cite{meza2023defcon, app-graceful1}. Graceful degradation may be independently employed by the application or the infrastructure, or through coordination between both, as shown in Figure~\ref{fig:compare}.

\parait{Application-level Graceful Degradation}: Past work introduced graceful degradation solutions for web servers~\cite{web-graceful1, web-graceful2}, mail services~\cite{mail-graceful1}, search engines~\cite{search-graceful1}, and storage systems~\cite{storage-graceful1,storage-graceful2,storage-graceful3}. Additionally, there exist solutions broadly applicable across applications~\cite{app-graceful1,app-graceful2}. Load shedding, or dropping a fraction of the load, is a common method employed at the application level~\cite{load-shedding1,load-shedding2,load-shedding3, cho2020overload, zhou2018overload, beyer2016site}. Brownout solutions~\cite{xu2019brownout, klein2014brownout}, on the other hand, allow the dimming of optional features at the application level. Microservice-based applications also leverage out-of-box circuit breakers~\cite{circuit-breaker} from commercial tools~\cite{hystrix, resilience4j,consul,istio} to degrade non-critical containers. These application-focused solutions perform degradations \textit{obliviously}, i.e., applications respond to capacity crunch scenarios without awareness of the extent of failures in the underlying infrastructure. 

\parait{Infrastructure-level Graceful Degradation}: In the infrastructure-only context, public clouds treat applications as blackboxes and rely solely on infrastructure-level signals for graceful degradation. For example, Kumbhare et al.~\cite{kumbhare2021prediction} infer the criticality of VMs by analyzing diurnal behaviors in access patterns and throttle non-user-facing ones. Similarly, other solutions~\cite{levy2020predictive, wu2012netpilot} choose potential mitigation actions (such as live migration) based on heuristics that are agnostic to application requirements. In addition, several infrastructure tools and techniques exist for application degradation, such as pod preemption based on pod priority in Kubernetes~\cite{podpriority}, and resilience guidebooks employed by cloud providers~\cite{google-dir-plan, aws-design-patterns, aws-resilience, composite-SLAs}. However, these solutions remain mostly unaware of application requirements. Moreover, current solutions do not support coordinated site-wide degradation policies.

\parait{Co-operative Graceful Degradation}: In cooperative degradation, cloud operators take mitigation actions with application awareness. Cooperative degradation offers faster response times and better resource efficiency since degradations are orchestrated at the cloud level with full visibility into infrastructure failures and application flexibilities. Recently, Meta introduced Defcon~\cite{meza2023defcon}, a resilience solution that employs cooperative graceful degradation in their first-party cloud, leveraging their visibility into both applications and the infrastructure. Defcon involves modification of Meta's own cloud applications to include \textit{knobs} that annotate program elements eligible for degradation. This allows the cluster manager to disable low-priority features during disaster scenarios. However, this approach requiring application-level modifications are not well-suited for public clouds.

\parab{Towards Practical Cooperative Graceful Degradation in Public Clouds:} The key goal of graceful degradation is to minimize the impact on application availability. A practical cooperative graceful degradation solution will not only consider application-level impact but will also leverage application awareness at the cloud level to improve response time and efficiency by \textit{prioritizing critical workloads over non-critical ones}. However, we face two key challenges in achieving this goal in public clouds today.

First, we need a standardized interface through which applications can clearly express their resilience requirements. An effective interface for cooperative graceful degradation in public clouds should (a) be general, easy to convey, and straightforward to adopt for a wide range of applications without requiring any application modifications, and (b) provide concise, actionable information for cloud operators. Critical functions may be identified internally or externally to the application. Defcon's approach is internal, with in-application knobs for degradation control. However, this approach is only suited for first-party clouds with complete application visibility and presents significant adoption barriers in public cloud environments. Hence, in public clouds, an interface that identifies critical components externally for black-box applications is essential.

Second, to enable automated resilience management, cloud operators need straightforward mechanisms to reallocate capacity from non-critical to critical workloads dynamically. While Defcon's knobs can help reduce server load, they lack explicit controls for dynamic capacity reallocation and introduce complexities in estimating resource savings. We need an appropriate level of granularity for resilience specifications that will enable seamless capacity reallocation by resilience management systems.

We make the case that containers offer an ideal abstraction to enable cooperative degradation in public clouds. In widely adopted microservice-based architectures, application functionality is decomposed into microservices, each deployed as an isolated container. These microservices, developed and deployed independently, are naturally suited to container-level degradation. Containerized degradation offers several advantages. First, it enables criticality specification at the container level without modifying or exposing application logic. Second, current container-based frameworks already support tagging at the container level; we can leverage this existing capability for resilience specification, thereby facilitating easy adoption. Finally, the resource savings from turning off containers are straightforward to estimate from container specifications, giving resilience management systems clear visibility into potential capacity reallocation gains. Building on these insights, we put forward the notion of diagonal scaling as a method to achieve effective, cooperative degradation in public clouds.

\color{black}

\vspace{-1mm}
\section{Diagonal Scaling}
\label{sec:ds}

\begin{figure}[t!]
	\centering
	\includegraphics[width=0.89\linewidth]{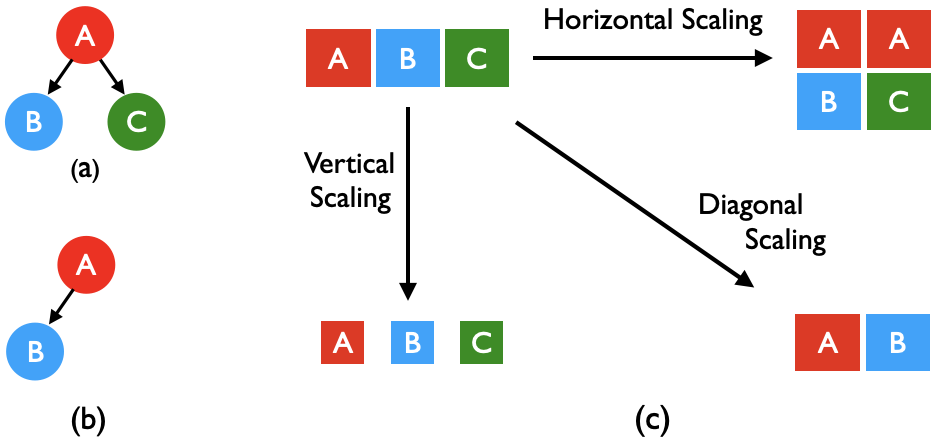}
	\caption{\small \saj{ \textbf{Diagonal Scaling.} (a) Original application dependency graph (DG) with 3 microservices. (b) Diagonally scaled DG with one microservice removed. (c) Comparison of horizontal, vertical, and diagonal scaling techniques.} }
	\label{fig:diagonal}
    \vspace{-3mm}
\end{figure}

We put forward \textit{diagonal scaling}, a graceful degradation technique at the container level that involves pruning applications by turning off microservices. This can help in improving the overall availability of ``critical'' services in the infrastructure during capacity crunch scenarios. Diagonal scaling is orthogonal to the notion of horizontal scaling (multiple parallel instantiations of containers) and vertical scaling (scaling up/down the resources allocated to containers) and may be employed alongside other scaling techniques. Figure~\ref{fig:diagonal} depicts a comparison of these scaling schemes.

To enable graceful degradation with diagonal scaling, we first need a mechanism for applications to indicate their preferences to the operator. 

\parab{Criticality Tags:} 
We introduce \textit{Criticality tags} as a simple yet expressive mechanism to indicate the importance or criticality of the microservice in an application. They capture criticality levels ($C_1$, $C_2$, $C_3$, etc.), with a lower number representing higher importance. We tag a container as high criticality (e.g., $C_1$) when it is key in driving the business of an application and a low criticality such as $C_5$ when it is “good-to-have.” For example, a document-editing application such as Overleaf may indicate chat as having lower criticality. By specifying a lower criticality tag, the application is agreeing that in case of a disaster, these microservices may be safely turned off. We can leverage existing tagging mechanisms in cluster management frameworks~\cite{kubernetes, tirmazi2020borg, tang2020twine, hindman2011mesos} to indicate criticality tags.

\subsection{Implications of Diagonal Scaling}

Diagonal scaling expands the resilience metrics space, thereby enhancing the granularity of resilience management. Resilience is typically measured using metrics such as the Recovery Time Objective (RTO), which specifies the maximum duration an application can tolerate being unavailable. Traditionally, resilience metrics are defined under the assumption that an application is either fully available or unavailable.

With diagonal scaling, an application can operate at multiple levels of availability. In addition to the fully "on" state, where all components are active, an application can serve a large number of user requests at multiple valid intermediate states, with only a carefully selected subset of components active. Thus, with diagonal scaling, RTO can be defined for various levels of operation. An application could define a stringent RTO for its critical functionality and a more lenient RTO for its auxiliary services. A broader range of resilience metrics provides operators with greater flexibility to maintain the availability of critical subservices during failures.

\subsection{Practical adoption}

We demonstrate diagonal scaling in a real-world application, Overleaf~\cite{overleaf}. We also discuss challenges with practical adoption and potential opportunities.

\parab{Real-World Application Demonstration}: Overleaf~\cite{overleaf} is a shared Latex editing environment. It comprises 14 microservices, such as spell-check and clsi. Users typically log in, open a project, and then perform edits and compiles on their Latex documents. Edits, which require low latencies, are implemented as web socket connections, whereas most other services are implemented as REST calls. Due to the decoupling of features in Overleaf, we observe that the application functions smoothly even when some non-critical microservices, such as chat and project tagging, are turned off. 
We first define a key business metric in Overleaf: edits made per second. We then evaluate the impact of turning microservices off on this key metric. We tag microservices that contribute highly to the metric as $C_1$ and microservices with low impact as $C_5$. We demonstrate that Overleaf can work seamlessly when $C_5$ microservices are turned off (\S\ref{sec:eval}).

\parab{Practical Challenges and Opportunities} 

\parait{\textbf{Rule-Based Criticality Tagging}}: We identify practical opportunities for rule-based criticality tagging.

\noindent
(i)\textit{ Microservices serving a single upstream caller}: We analyze the microservice dataset from Alibaba [3]. We derive dependency graphs of 18 applications from this dataset, consisting of over 20 million call graphs following the dependency mining methodology [4]. We find that ~74\% of microservices in the top 4 applications, and 82\% across all 18 applications are invoked by a single upstream microservice. These "single-upstream" stub microservices can be safely degraded if marked as low criticality by the upstream caller without causing any cascading failures. A similar analysis on Meta’s microservice deployment fleet showed that ~60\% of microservices are ML-inference based, serving a single upstream microservice~\cite{huye2023lifting}. We posit that these stub microservices are suitable candidates for criticality tagging.

\noindent
\textit{(ii) Frequency-based criticality tagging}:  Among 18 applications in Alibaba traces, we find that most requests are served by four applications. Using an LP, we determine for each application the minimal set of microservices required to serve the maximal number of requests (refer Appendix~\ref{appendix:alibaba}). We find that in the most popular application, which serves over 1.3 million requests and contains ~3000 microservices, more than 80\% of requests can be served by enabling only 3\% microservices (90 microservices). Other applications also exhibit similar behavior, indicating a large skew. We argue that a frequency-based classification of microservices can be useful for applications to identify which microservices are more critical than other microservices.

\parait{\textbf{Automated Criticality Tagging and Testing}}: While manual tagging is feasible in small applications, it can be tedious for large ones built by several teams. We envision application developers leveraging their system logs to infer criticalities using learning-based schemes. While data-driven insights are helpful, application developers may need to run additional tests to assess the effectiveness of their criticality tagging. Furthermore, application developers may need to override and tag known high-criticality low-frequency microservices manually. To assess criticality tagging effectiveness, we also build a chaos-testing framework (\S\ref{sec:implement}).

\parait{\textbf{Support for Flexible Adoption of Tagging}}: Not all applications may be amenable to container-level degradation. For example, when a single microservice contains both critical and non-critical functionalities. Therefore, our system design does not require all applications or all microservices within an application to be assigned criticality tags. In such cases, all untagged microservices will be deemed to be of the highest criticality level. 

Our goal is to leverage the benefits of cooperative degradation on the fraction of existing applications which can diagonally scale. Nonetheless, the advent of modern cloud development architectures such as Service Weaver~\cite{ghemawat2023towards} shows promise in that they propose a container runtime framework that will offload how application binaries are packed and shipped to the container runtime. In such cases, developers can specify the criticality on the code-interface level which can then be leveraged by the container-runtime policy to separate critical and non-critical containers for improving the resilience of the application.

\color{black}

\vspace{-2mm}
\section{System Design}
\label{sec:design}
We present the design of Phoenix that performs application-aware resilience management at data center scale. Phoenix's key goal is to satisfy the resilience objectives of applications and operators maximally in the event of large-scale failures. Phoenix takes into account the following considerations:

\begin{itemize}[leftmargin=0cm,nolistsep,label={}]
    \item \textbf{(R1) Application Requirements:} Within an application, microservices should be enabled based on the order of criticality. When dependency graphs are available, microservices are ordered based on both dependencies and criticality.
    \item \textbf{(R2) Operator Objectives}: While criticality tags enable to prioritization of containers \textit{within} an application, the operator objectives determine the resource allocation \textit{across} applications. Phoenix should support a variety of operator objectives (e.g., fairness, revenue maximization). 
    \item \textbf{(R3) Resource Efficiency}: Phoenix should efficiently pack microservices within the available cluster capacity.
    \item \textbf{(R4) Fast Response Time}: Phoenix should be responsive to disasters in the order of seconds at the data center scale.
    \item \textbf{(R5) Broad Deployability}: Phoenix should be broadly deployable across applications and data center environments. It should be able to handle clusters with a mix of applications that may/may not be diagonally scalable and may/may not have dependency graphs.
\end{itemize}

We design the resilience management system as an independent layer that interacts with cluster schedulers. Phoenix constantly tracks the cluster state and, during a failure event, generates a new target state based on operator and application goals, which is conveyed to the cluster scheduler. This decoupling of resilience management from the cluster scheduler enables Phoenix to be easily portable across cluster schedulers. It also allows us to maintain a simpler design at each of the individual layers while also supporting independent evolution at both layers. Finally, this separation will enable future extensions of Phoenix to handle resilience across the cluster scheduler, the network controller, and the storage controller. 

We first develop a Linear Program (LP) that adheres to the above design requirements.

\parab{Linear Program Formulation}: We use the following variables in the LP. The Boolean variable $x_{ij}$ represents whether a microservice $j$ in application $i$ is activated. The Boolean variable $y_{ijk}$ represents whether a microservice $j$ in application $i$ is placed on server $k$.  $C(m_j)$ denotes the criticality level of microservice $m_j$ --- lower numbers, $C_1$, imply higher criticality. $R_{ij}$ represents the resource requirement of microservice $m_j$ belonging to application $app_i$ and $pred_i(m_j)$ refers to its predecessors in the dependency graph. $S_k$ denotes the capacity of the server $k$. $F(x_{ij})$ represents the global ranking function, which could be fair share, revenue, etc.
The Linear Program (LP) formulation that captures Phoenix requirements can be written as follows:

\vspace{-3mm}
\begin{align*}
    &Maximize \; \sum_{i} \sum_{j} F(x_{ij}) 
\end{align*}
\begin{align}
x_{ij} \geq x_{ik} \;|\;\forall m_j , &m_k \;\epsilon \;app_i \; with \; C(m_k) > C(m_j)\\
\sum_{j\; \epsilon \; pred_i(m_k)}x_{ij} \geq x_{ik}\;&|\; \forall app_{i}, \; \forall m_k \;\epsilon \; app_i \\
\sum_{k}y_{ijk} = x_{ij} \;&|\; \forall app_{i}, \;\forall m_j \;\epsilon \; app_i\\
\sum_{i}\sum_{j}R_{ij}*y_{ijk} < S_k \;&| \;\forall S_{k}, \;\forall app_i, \;\forall m_j \;\epsilon \; app_i
\end{align}

Eq. (1) specifies criticality constraints within an application, where nodes are activated in the order of criticality. For example, $C_1$ microservices are activated before activating $C_2$. Note that this constraint applies within each application only and not across applications. Eq. (2) imposes topological constraints wherein for each node $m_k$, at least one predecessor of $m_k$ must be activated. This implies that in the application dependency graph, there is at least one activated path leading to every activated node. This constraint is optional when the service dependency graph for an application is missing. Eq. (3) specifies that a microservice $m_j$ of $app_i$ must be placed in at most one server. Eq. (4) imposes that the total assigned load on a server $k$, is less than its capacity, $S_k$.

\parab{Global Objectives}: Globally, the operator objective, $F$, determines the resource allocation across applications. The operator has the flexibility to define any monotonically increasing function 
$F$ as an objective. 

We consider two candidates for $F$: 1) Fairness-based and 2) Revenue-based. In the fairness-based objective, the goal is to distribute $R$ units of resources among $n$ applications such that each application receives a fair share $R/n$. If an application’s demand is larger than $R/n$, it is allocated at least (but possibly more than) this share. However, if the job’s demand is smaller, the excess resources may be allocated to other jobs to improve overall resource usage. We formulate a water-filling~\cite{kumar2015bwe, fair-share-tut, radunovic2007unified} based fairness objective to implement this. In the revenue-based objective, the LP prioritizes applications with a willingness to pay a higher price per unit of resource. We refer to these two LP formulations as LPFair and LPCost, respectively. The mathematical formulations of these LPs can be found in Appendix~\ref{appendix:lp}.

Our LP formulation is designed to be generic and adaptable, allowing the addition of various constraints relevant to cloud operators. Examples of such constraints include limiting the migration of microservices from unaffected nodes or adhering to per-node microservice limits imposed by underlying cluster schedulers. However, as we demonstrate in \S~\ref{sec:eval}, LP-based solutions scale poorly to today's cluster sizes. Hence, we use the LP as a guide to design the Phoenix system.

\begin{figure}[t!]
	\centering
	\includegraphics[width=0.9\linewidth]{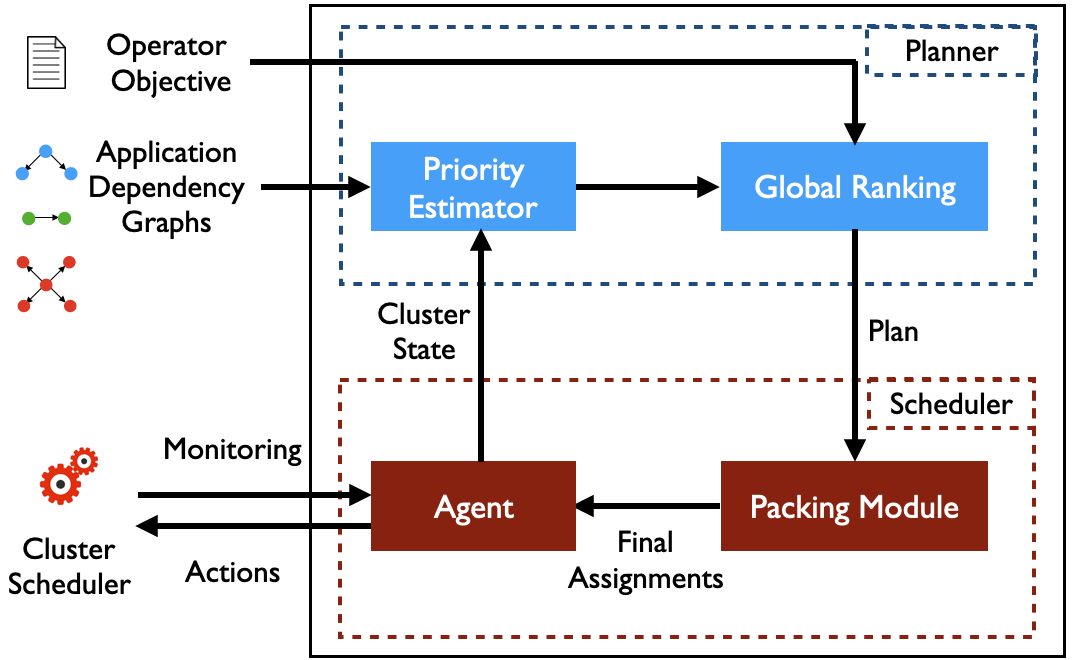}
	\caption{\small Phoenix System Diagram} 
	\label{fig:sys}
 \vspace{-12pt}
\end{figure}

\parab{Phoenix}: The Phoenix system (Figure \ref{fig:sys}) has two key components: a Planner and a Scheduler. The planner generates a plan for allocating resources to microservices during a failure scenario. The planner takes as input microservice-level information of active applications in a standardized format, including criticality tags and resource allocation before failure. The planner then generates a subset of microservices to be enabled by considering the aggregate resource availability in the cluster. The scheduler is responsible for generating a mapping from microservices to nodes in the cluster and enforcing this plan in the correct sequence by interfacing with the cluster scheduler.

\subsection{Phoenix Planner}
\label{sec:planner}

The planner consists of two sub-modules (Alg.~\ref{code:cap}). The \textit{Priority Estimator} generates an ordered list of microservices per application based on application requirements: criticality tags and (optionally) application dependency graphs. The \textit{Global Ranking} submodule generates a globally ordered list of microservices based on the operator objective and application-level ranking generated by the priority estimator.

\begin{algorithm}[t]
\SetKwInput{KwInput}{Input}                %
\SetKwInput{KwOutput}{Output}              %
\DontPrintSemicolon
  \SetKwFunction{FMain}{Main}
  \SetKwFunction{FGlob}{GetGlobalRank}
  \SetKwFunction{FGlobGlob}{CustomKey}
  \SetKwFunction{FSub}{PriorityEstimator}
  \SetKwFunction{FSubSub}{DFS}
  \SetKwFunction{FFairShare}{GetFairShare}
  \SetKwFunction{FSort}{Sort}
  \SetKwFunction{FSum}{Sum}
  \SetKwFunction{FBidding}{GetBiddingRank}
  \SetKwFunction{FPriQ}{InitPriQ}
 
\SetKwProg{Fn}{Function}{:}{\KwRet}
\Fn{\FMain}{
    AppRank = \FSub{apps, tags}\;
    GlobalRank = \FSort(AppRank,ClusterObj)\;
    \KwRet GlobalRank
}

\SetKwProg{Fn}{def}{:}{}
\Fn{\FSub{$apps$, $tags$}}{
    \Fn{\FSubSub{$node$}}{
            \lIf{$node \in visited$}{\KwRet}
            AppRank.append((app.id, node.id))\;
            \ForEach{$child \in node.children$}{
                \lIf{$tags(child) \geq tags(node)$}{\FSubSub{$child$}}
                \lElse{$Q.insert(child)$}}    
    }

    \ForEach{$app \in apps$}{
        \If{$app.G \neq NULL$}{
        Q = \FPriQ{G.src\_nodes, key=tags}\;
        \While{$len(Q) \neq 0$}{
        \FSubSub{$Q.pop()$}}}
        \Else{
        \ForEach{$node \in sorted(app.nodes)$}{
            AppRank.append((app.id, node.id))
        }}
        }
    \KwRet AppRank}

\SetKwProg{Fn}{def}{:}{}
\Fn{\FGlob{AppRank,Obj}}{
Q = \FPriQ{[root for app in AppRank],  key=Obj}\;
\While{$len(Q) \neq 0$}{
(appID, ms) = Q.pop(0)\;
R = R - ms.resources\;
\If{$R \geq 0$}{
GlobalRank.append(appID, msID)\;
Q.insert(AppRank[appID][ms.idx+1])}
}
\lElse{break}
\KwRet GlobalRank
}
    
\caption{Criticality-Aware Planning Algorithm}
\label{code:cap}
\end{algorithm}

\parab{Priority Estimator:} This module determines the relative priority of containers within an application based on their criticality and dependencies. The Priority Estimator (in Algorithm~\ref{code:cap}) takes as input an `app` object representing the list of containers and criticality tags and optionally dependency graphs (DGs) when available. Its output is an ordered list, $AppRank$, which denotes the priority order in which microservices need to be activated. Note that this ordering is within each application.

When DGs are unavailable (line 13), containers are ordered based on criticality from highest to lowest (lines 18 and 19). For applications with DGs, we populate a priority queue, $Q$, with source nodes i.e., entry microservices with no inbound edges (line 14). In lines 14-19, we traverse the graph based on a combination of two factors---criticality (from high to low) and topological ordering (from root to leaves). The pre-order graph traversal ensures that the ordering of nodes is topology-aware such that no microservices are activated without at least one predecessor (satisfying Eq. (2) in the ILP formulation). Using criticality as the key while popping nodes from $Q$ ensures that our orderings are criticality-aware, satisfying Eq. (1) in the ILP formulation. We avoid redundant computation by maintaining a visited set (not shown in the algorithm) to reduce the time complexity for each graph to be similar to a DFS/BFS traversal in $O(V+E)$ time.

\parab{Global Ranking:} This module leverages cluster operator objectives to obtain a global ordering of microservices across all applications. As shown in line 21 in Algorithm~\ref{code:cap}, this module takes as inputs the operator objective, $Obj$, and the ordered lists of containers per application generated by the priority estimator, $AppRank$. The output is $GlobalRank$, an ordered list of containers across all applications. \kapil{$Obj$ in Algorithm~\ref{code:cap} is an operator-defined Python method that takes as input the list of applications, container resources, and current assignment state, and outputs a score. This scoring function captures the objective by which cloud operators prioritize and rank microservices \textit{across applications}, such as fairness or revenue maximization.}

We use a priority queue, $Q$, to track the containers sorted based on the operator objective. $Q$ is initialized with the first node in every application's priority list in line 22. In each step, the container with the highest value based on the operator's objective is popped from the priority queue and added to $GlobalRank$ in lines 24-28. The resource usage of the container is deducted from the available capacity in line 25. The next node in the corresponding application's priority list is added to the priority queue. This process continues until all containers are visited.

Phoenix planner supports a broad range of operator objectives. Following the LP formulation, we implement two operator objectives: \textit{(i) Cost-Based:} Containers that generate higher revenue are prioritized. The key in the cost-based global ranking is the price per unit resource.  \textit{(ii) Fairness-Based}: Those containers are allocated resources in every round whose resulting deviation is least from the precomputed fair-share.

\subsection{Phoenix Scheduler}

The scheduler is responsible for mapping containers to servers based on the ordered list generated by the planner. This formulation is a variation of the well-known bin-packing problem~\cite{binpacking} and is known to be NP-hard. Hence, we design a criticality-aware scheduling heuristic. The scheduler has two modules: the packing module, which generates the mapping, and an agent that executes it. 

\parab{Packing module:} The packing module is responsible for mapping the containers to servers based on the ordered list generated by the planner. Note that this module performs all operations on a copy of the cluster state and does not enforce them on the cluster. The final execution is deferred to the Agent. The detailed pseudocode is given in Algorithm~\ref{hscheduler} in Appendix~\ref{sched:code}. We highlight the key steps here.

When a cluster experiences partial failures, a fraction of containers in the planner list may be already running. The packing module iterates over the ordered list generated by the planner. If the next container to be scheduled is already running, it can continue on the same server. If the container to be activated has failed, it must be rescheduled on an active server. The scheduling heuristic first sorts the active servers in the decreasing order of available capacity. If the resource requirements of the container can be accommodated without migrating any of the other active containers, it is assigned to the server that has the smallest available capacity larger than the required resources, i.e., the best-fit strategy~\cite{bestfit}. 

If the container cannot be accommodated on the existing servers based on the available capacity, the heuristic will proceed to find a migration strategy. It identifies a source server from which containers can be migrated based on available server capacity and the size and count of containers currently active on the server. Smaller containers are more likely to be accommodated by other servers. Hence, servers with large available capacity and a large number of small currently active containers are preferred. The heuristic then migrates the containers on the server to other active servers. If the heuristic fails to identify a target server for a container under both best-fit and migration strategies, it proceeds to delete the currently active services in the reverse order from the planner's list. The lowest-priority containers are deleted first. After each deletion, the heuristic attempts best-fit and migration strategies again to find a suitable mapping.

\parab{Agent:} The agent continuously monitors the cluster state, and when a failure event is detected, reports it to the planner. The agent is also responsible for executing the list of tasks determined by the Phoenix scheduler on the cluster scheduler. At a high level, the agent performs three tasks: deleting non-critical containers, migrating already running containers, and restarting containers impacted by a failure. We use cluster scheduler API to execute these tasks. Additional subroutines performed with all three tasks include draining traffic, scaling up/down the containers, and reconfiguring iptables, detailed in the implementation section.

\vspace{-2mm}
\section{Implementation}
\label{sec:implement}

The Phoenix Controller is written in Python and is available as open source~\cite{phoenix}. We test Phoenix Controller with Kubernetes cluster scheduler. Note that our design can also work with other cluster schedulers~\cite{tang2020twine, hindman2011mesos, vavilapalli2013apache} with minor modifications. The Phoenix Agent monitors the cluster state at 15-second granularity. This is a tunable parameter. We chose 15 seconds to maintain a low response time while ensuring the Kubernetes cluster is not overwhelmed. The Phoenix controller maintains the deployment information and associated criticality tags. It also maintains application dependency graphs as NetworkX DiGraph objects \cite{networkx}. In the packing module, we employ a tree-based data structure, Python’s Sorted Lists \cite{sortedlist}, to perform insert, search, and delete operations faster than linear time.

\parab{Fault Tolerance}: Phoenix is a lightweight module that can tolerate unplanned failures. While Phoenix maintains the information of criticality tags and DGs in memory, these inputs are also persisted on a storage service that can be fetched on-demand. When failures occur, leading to a sudden crash for Phoenix, it can simply restart on a healthy node, pull the inputs from a persistent store, and resume operation.

\parab{Chaos Testing Service}: Since an application development life-cycle consists of regular rollouts, rollbacks, version updates, etc., we also build a managed chaos testing suite in Phoenix to improve developer productivity. This testing service injects failures to verify the correct behavior of an application under the specified criticality tags. It takes as input the application's deployment files (such as YAML, and TOML), an end-to-end load-generator (such as Locust or wrk2), and a utility function. The utility function computes a score on the logs generated by the load generator to measure the quality of the application's outputs. Before pushing deployments to production, this service can conduct tests at different degrees of failure and report the results to developers.

\parab{Partial Tagging}: Phoenix can operate when only a subset of applications are diagonal scaling compliant. We achieve this by using labels on namespaces, tagging only the subscribed applications as "phoenix=enabled". Furthermore, applications can also partially tag their degradable containers. Phoenix, by default, assumes the criticality to be highest when no criticality tags on deployments are specified.  

\parab{Stateless Workloads}: As noted in \S~\ref{sec:intro}, Phoenix currently only supports diagonal scaling on stateless services. Nonetheless, stateless workloads comprise more than 60\% of large data center machine usage, as reported by real-world data centers~\cite{lee2021shard}. We expect significant benefits at current levels since Phoenix can provide benefits even when a fraction of applications are not diagonal scaling compliant.

\parab{Diagonal Scaling Practical Experience}: We now discuss our process of adapting HotelReservation (HR)---a microservice based reservation application from DeathStarBench~\cite{gan2019open}---to support diagonal scaling. Since Overleaf is diagonal-scaling compliant, we first inspect its code to apply our learning to HR. Our inspection of Overleaf’s code showed that it is crash-proof. When a functionality, such as spell-check or chat, is turned off, Overleaf can continue serving requests with reduced functionality, thereby remaining compliant with diagonal scaling requirements. This resilience is achieved through several application-level measures, including error handlers that wrap downstream calls to silently handle failures and allow binaries to execute successfully.

In contrast, HR is primarily designed as a demonstration application for research purposes and, therefore, lacks robust error-handling mechanisms. Although HR, like Overleaf, separates non-critical features into distinct microservices, it is not entirely crash-proof; disabling certain non-essential microservices can still result in user-visible failures. For example, the initialization of the front-end server depends on the availability and connectivity of downstream microservices like search, profile, user, reservation, and recommendation. To enable diagonal scaling, the front end should remain stable even if a low-criticality service, such as recommendation, is turned off. To address this, we implement error-handling logic to prevent request crashes when a downstream microservice, such as the user microservice, is unavailable.

\vspace{-1mm}

\section{Evaluation}
\label{sec:eval}

We evaluate Phoenix to answer the following questions:
\begin{itemize}[leftmargin=*,nolistsep]
\item Can Phoenix's cooperative degradation improve the availability of cloud applications?
\item Can Phoenix perform well at scale in clusters with over 100,000 servers and across real-world application dependency graphs with thousands of microservices?

\item What are the performance implications of Phoenix at both the cloud operator level (time to mitigate) and at the application level (performance degradation)? 
\end{itemize}

We conduct our experiments across two distinct environments: (1) CloudLab, where we deploy two microservice-based workloads on a cluster with 200 CPU cores; and (2) AdaptLab, our benchmarking platform, to simulate sub-data-center failures in realistic large-scale cloud environments up to 100,000 nodes.

\begin{figure*}
\begin{minipage}{0.33\textwidth}
    \begin{tabular}{c|c}
       \textbf{Application}  & \textbf{Metric} \\
       \hline
        Overleaf0 & document-edits\\
        Overleaf1 & versions\\
        Overleaf2 & downloads\\
        \hline
        HR0 & search \\
        HR1 & reserve
        
    \end{tabular}
    \caption{\small Resilience objective of individual applications. For example, we say Overleaf0's resilience goal is satisfied if the served requests-per-second (RPS) of document-edits remains unaffected.}
    \label{tab:resilience-per-app}
        
\end{minipage} \hfill
\begin{minipage}{0.63\textwidth}
    \begin{subfigure}{0.45\linewidth}
    \includegraphics[width=\linewidth]{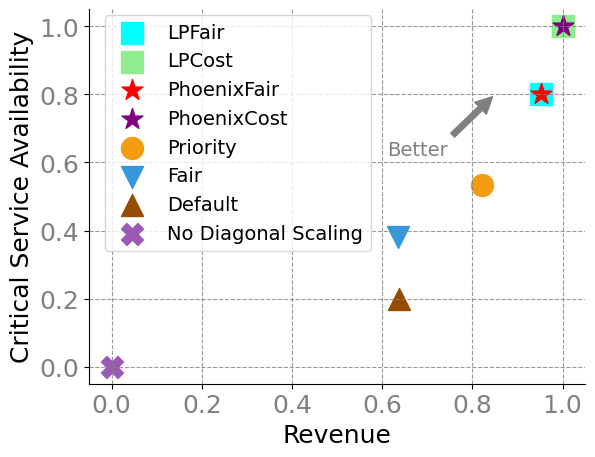}
    \caption{Revenue}
    \end{subfigure} \hfill
    \begin{subfigure}{0.45\linewidth}
    \includegraphics[width=\linewidth]{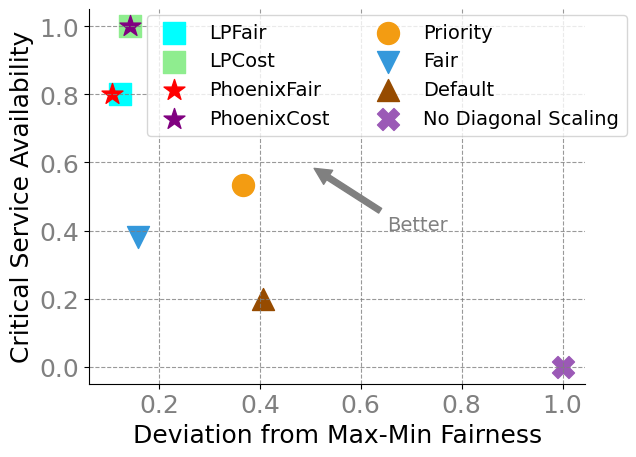}
    \caption{Deviation from Fairness}
    \end{subfigure}
        \caption{\small Resilience schemes evaluated on a Kubernetes cluster with cluster capacity reduced to ~42\%. Critical service availability across microservice applications (with heterogeneous goals based on Table \ref{tab:resilience-per-app}) is shown. The x-axis shows the operator objectives.}
        \label{fig:scatter}
\end{minipage}
\end{figure*}

\parab{Baselines}: We evaluate two variants of Phoenix---Phoenix-Fair (operator objective is max-min fairness) and PhoenixCost (operator objective is to maximize revenue) as discussed in~\S\ref{sec:design}---with their corresponding LP formulations, LPCost and LPFair. We also compare Phoenix with two non-cooperative degradation baselines. We implement a fairness-based resource redistribution scheme, \textit{Fair}, which does not take into account criticality tags. In the second non-cooperative baseline scheme, \textit{Priority}, applications expose criticality tags, but the operator does not enforce any per-application quotas. The baseline \textit{Default} represents the default behavior of Kubernetes without information on criticality tags or cluster operator objectives.

\parab{Operator Metrics}: We report operator metrics such as cluster utilization, revenue, and resource fairness at varying failure rates. Revenue is computed based on whether a microservice is activated or not when failures strike. Fairness is measured as the deviation from max-min fairness. We decompose the fairness deviation measure into two parts: positive deviation (using more resources than fair share) and negative deviation (using fewer resources than fair share).

\subsection{Phoenix with Kubernetes on CloudLab}

We deploy Overleaf~\cite{overleaf} and the Hotel Reservation (HR) system from DeathStarBench~\cite{gan2019open} on a CloudLab cluster with 200 CPUs, using d710 machines equipped with 64-bit Intel Quad Core Xeon E5530 processors and running Ubuntu 22.04 LTS, to simulate a real-world multi-tenant environment. Overleaf inherently supports diagonal scaling (as detailed in \S~\ref{sec:ds}), whereas HR requires minimal modifications to achieve diagonal scaling compliance (outlined in \S~\ref{sec:implement}).

\parab{Experimental Setup}: We run five instances across the two applications, as shown in Table \ref{tab:resilience-per-app}. For load generation, we use publicly available load generators for Overleaf~\cite{thalheim2017sieve, tato2018sharelatex} and wrk2 ~\cite{gan2019open} for HR. For each instance, we tweak the parameters so each application's resource distribution across containers is different. For example, different levels of edits, spell-checks, versioning, etc. We determine the resource requirements of containers by running their respective load generators. 

\parab{Criticality Tagging}: Following chaos engineering methodology~\cite{chaosprinciples, netflixchaosmetric}, we first define the steady state for each application, representing the application’s "healthy" operational behavior. Any disruption to this steady state indicates a failure to meet the application’s critical service goal. For each application, we designate one primary service, listed in Table \ref{tab:resilience-per-app}, as the most critical; this service's throughput defines the application's steady state. We label the microservices supporting this critical service as $C_1$ for each application. All other microservices are assigned lower criticality levels.

\parab{Application Metrics}: We define an application's \textit{critical service availability goal} as met when the requests per second (RPS) of the critical service are retained after a failure event and unmet otherwise. We assume stateful workloads such as MongoDB~\cite{mongo} are running on a separate stateful cluster, as is standard practice adopted by cloud applications today~\cite{luo2021characterizing}, running stateful and stateless workloads in separate clusters. To show the impact of degrading low-criticality features, we adopt the approach proposed in Fox et. al, 1999~\cite{fox1999harvest} of harvest and yield by assigning a utility (harvest) to each use-case. We augment the load generator to compute a utility. Each microservice is assigned a utility value that aligns with its criticality. The utility of a service is the sum of the utilities of the component microservices. 

\parab{Experimental Setup}: In all CloudLab experiments described below, we report results by reducing the cluster capacity to 42\% (i.e., the breaking point, going below which violates the minimum resource required for maintaining critical service, in Appendix~\ref{appendix:cloudlab-resource-dist}). The y-axis reports the critical service availability; it takes a value of 1 when all $C_1$ microservices of the critical service listed in Figure \ref{tab:resilience-per-app} are active and 0 otherwise. The x-axis shows two operator objectives: revenue in (a) and fair share deviation in (b).

\parab{Cooperative Degradation allows applications to withstand failures in-place.}The $\times$ marker (in purple) in Figures~\ref{fig:scatter} (a) and (b) represents no diagonal scaling, i.e., when applications cannot adapt to a resource crunch to resume operations. On the other hand, all other schemes have non-zero service availability and continue to operate at reduced capacity.

 \begin{figure*}[h]
    \centering
    \hspace*{-0.75cm}
    \begin{tabular}{ccc}
    \includegraphics[width=0.31\textwidth]{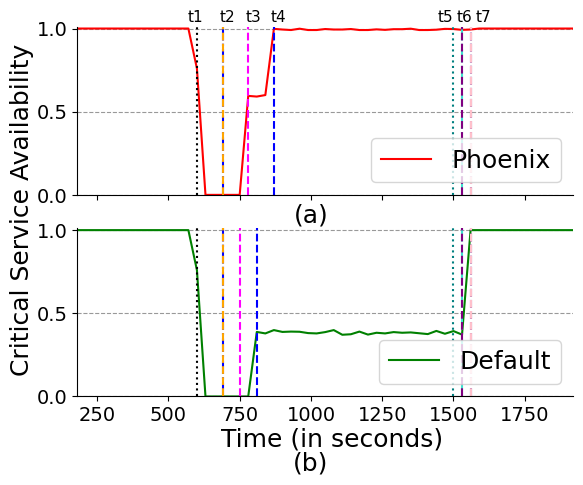} 
    &
    \includegraphics[width=0.31\textwidth]{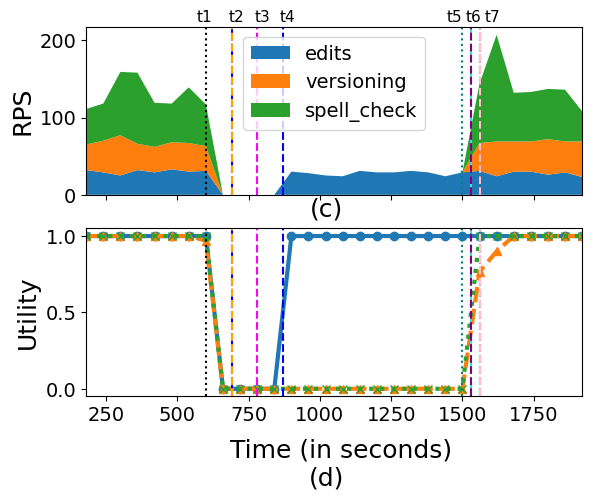} 
    &
    \includegraphics[width=0.31\textwidth]{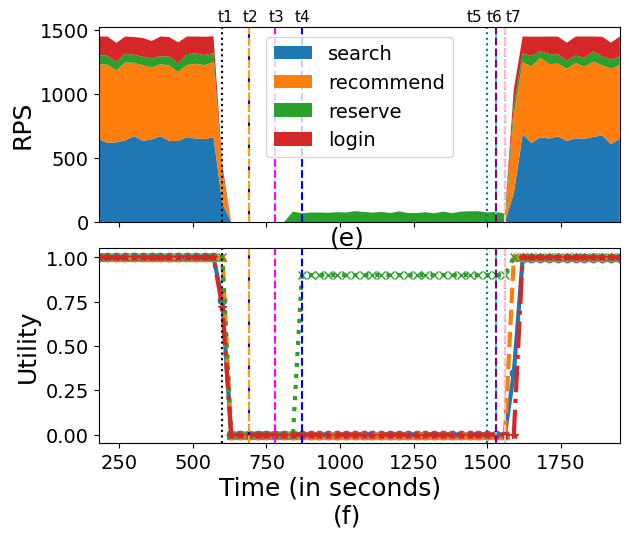} 
    \\
    \end{tabular}
    \caption{\small Diagonally scaling a multi-tenant cluster with microservice application instances using Phoenix. \textbf{(a)} and \textbf{(b)} show the benefits of Phoenix over Default when cluster capacity reduces to 40\%. \textbf{(c)} and \textbf{(e)} demonstrate diagonal scaling on Overleaf and HR, respectively, where critical service throughput (requests per second) is retained while non-critical services are turned off during resource-crunch scenarios. \textbf{(d)} and \textbf{(f)} show end-user utility degradation of different services under diagonal scaling. (f) demonstrates the degradation of optional yet "good-to-have" features as end-user utility drops to 0.8 for ``reserve''.}
    \label{fig:phoenix-cloudlab}
    \vspace{-1mm}
\end{figure*}

\parab{Phoenix has broad applicability both across applications and cloud environments.} Figures~\ref{fig:scatter} (a) and (b) show Phoenix's ability to maximize different operator objectives: revenue and fairness, respectively. Phoenix achieves superior performance under both objectives.

\parab{Phoenix performs targeted recovery of critical services.} We now report qualitative results by showing a real-world run. In this experiment, we measure the performance of PhoenixCost against the Kubernetes Default mechanism. To emulate actual failure events in the cloud, we stop the Kubelet process~\cite{kubelet} on the failed nodes and restart it after 10 minutes. We use the same detection mechanism, outlined in ~\S\ref{sec:implement} for both schemes. We mark events using time-markers at the top of the plot. Figures \ref{fig:phoenix-cloudlab} (a) and (b) show critical service availability of two runs with Phoenix and Default.

When a failure occurs (t1), Phoenix detects it after ~100 seconds and prepares a plan almost instantly (t2). At t2, the Phoenix agent then starts issuing the commands to the Kubernetes cluster to reach the target state (t3). The agent marks the cluster state as recovered when the desired state is reached (t4). The time elapsed between executing action (t3) and completion (t4) can vary depending on the pod deletion and startup times. At ~1500s, the nodes come back online (t5). Notice that Default only resumes its operation once all nodes are recovered at 1500s mark. Phoenix's target-driven recovery allows all 5/5 applications to retain critical service availability, whereas Default is able to satisfy only 2/5 (40\%).

From the same run in \ref{fig:phoenix-cloudlab} (a), we zoom in on two applications, Overleaf0 and HR1, and report requests per second served for each of the services. Figure \ref{fig:phoenix-cloudlab} (c) shows the throughput as requests per second of three request types, namely edits, spell-check, and versioning, plotted as a stacked chart for Overleaf0. The whitespace (between ~600s and ~900s, t1 - t4) in (c) represents application downtime due to frontend failure. We observe that the throughput of edit requests, the critical service for Overleaf0 (Table~\ref{tab:resilience-per-app}), recovers in under 4 minutes. 10 minutes later (at the ~1500s mark), when failed nodes recover, Phoenix can instantly detect the capacity increase and spawn non-critical services. The sharp rise in the spell\_check service immediately following the recovery is due to the pending edits during service downtime.

  \begin{figure*}[h]
    \centering
    \hspace*{-0.75cm}
    \begin{tabular}{ccc}
    \includegraphics[width=0.31\textwidth]{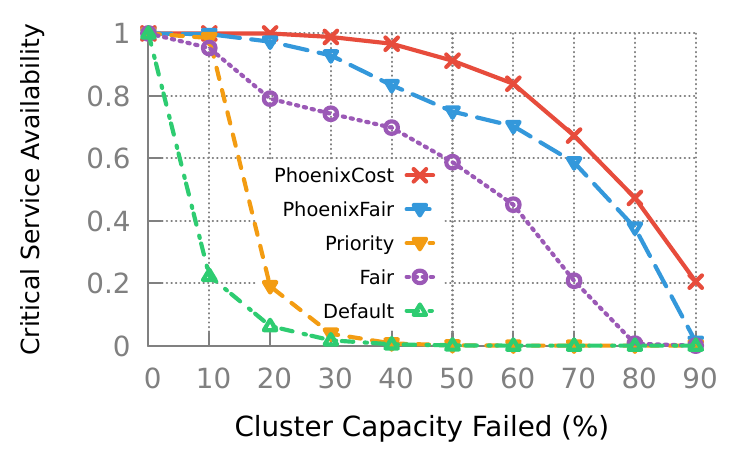} &
    \includegraphics[width=0.31\textwidth]{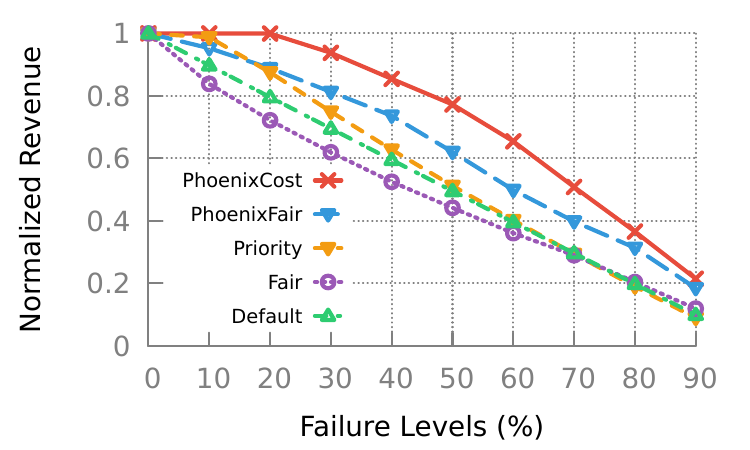} &
    \includegraphics[width=5cm,height=3.2cm]{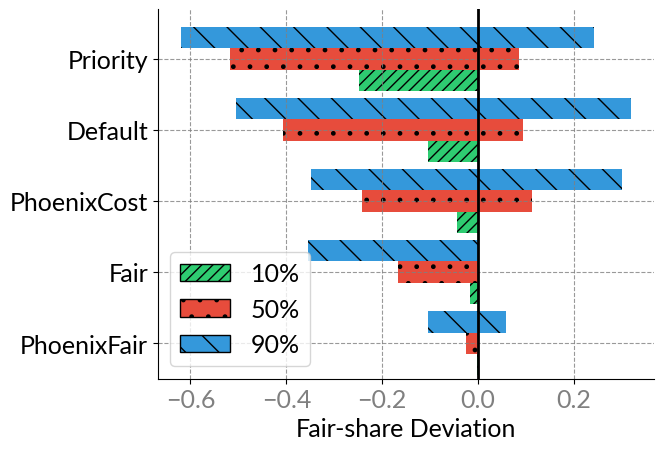} \\
    \end{tabular}
    \caption{\small Resilience schemes evaluated on AdaptLab using Alibaba traces, Service-Level-P90 criticality tagging scheme, and Calls-Per-Minute (CPM) based resource assignment scheme in a 100,000-node cluster. (a) Aggregate critical service availability across applications at different capacity failure scenarios shows that PhoenixFair and PhoenixCost activate more critical services consistently. (b) Normalized revenue shows that PhoenixCost maximizes revenue. (c) Deviation from fair-share shows that PhoenixFair has the least deviation. }
    \label{fig:freq_p90}
\end{figure*}

Figure~\ref{fig:phoenix-cloudlab} (d) reports the end-user's utility. Utility falls for versioning and spell-check when nodes fail, while edits maintain high utility. As noted above, we instrument load generator scripts to ascribe a utility score for each successful request. Utility is 0 when a request fails. Without any code modifications in Overleaf, Phoenix degrades non-critical services, such as spell-check and versioning, to ensure fast recovery of Overleaf's critical services.

\parab{Pruning call-graphs for partial utility.} We provide another example where a service can continue serving requests in a degraded mode by dropping optional yet "good-to-have" features. Figure~\ref{fig:phoenix-cloudlab}(e) shows how non-critical services are turned off by Phoenix while ensuring the throughput of ``reserve'' is maintained. Furthermore, in Figure~\ref{fig:phoenix-cloudlab}(f), we observe that the ``reserve'' utility was decreased to 0.8, showing that the end-user utility of reservation drops. This is a result of turning off a non-critical downstream call to the user, allowing reservations to be made as a guest. When the nodes recover, the ``reserve'' utility returns to 1. Note that when we partially prune a service, the performance may degrade due to timeouts. However, we do not observe any performance degradation in our P95 latency measurement (Appendix \S~\ref{appendix:latency-impact}).

\subsection{AdaptLab Phoenix Evaluation}

We develop AdaptLab, a resilience benchmarking platform to emulate failures in large-scale clusters. We evaluate Phoenix and baselines at scale using real-world traces from Alibaba clusters~\cite{luo2021characterizing} on AdaptLab. The Alibaba dataset contains over twenty million call graphs collected over a seven-day period~\cite{luo2021characterizing}. Using the methodology from Luo et al., 2022~\cite{luo2022depth}, we derive 18 application dependency graphs of varying sizes (ranging from 10 to 3,000 microservices). Since these traces do not include specific CPU/memory usage data for each microservice or information on criticality assignment, we experiment with two resource allocation models and two criticality assignment models within this environment.

\parab{Resource Assignment}: We test two realistic resource models to approximate the resource requirements of each microservice: (i) resources as a function of calls-per-minute, proposed by another study from Alibaba on the same dataset~\cite{luo2022power}, and (ii) resources sampled from a long-tailed distribution model as specified in Azure Bin-packing traces~\cite{azurepublicdata}.

\parab{Criticality Tagging}: We develop two schemes for criticality tagging in AdaptLab for Alibaba's application graphs: (i) service-level tagging and (ii) frequency-based tagging. Service, in this context, refers to a set of microservices that together offer a useful functionality. In \textit{service-level tagging}, we identify the most frequently invoked services and assign all the component microservices as $C_1$. In \textit{frequency-based tagging}, we use a linear program to find the top microservices that can serve specified target percentile requests, as reported in Appendix \S\ref{appendix:alibaba}. We generate both service-level tagging and frequency-based tagging at $50^{th}$ and $90^{th}$ percentile, denoted as P50 and P90, respectively. In addition, in all schemes, we tag a tiny fraction of infrequently invoked services that are randomly chosen as highly critical. This is to account for critical background services that are infrequent, such as garbage collection routines. 

\parab{Application Metrics:} We define an application's \textit{critical service availability goals} as met when ``all'' $C_1$ microservices are running.

\begin{figure*}[h]
    \centering
    \begin{tabular}{ccc}
    \includegraphics[width=0.4\textwidth]{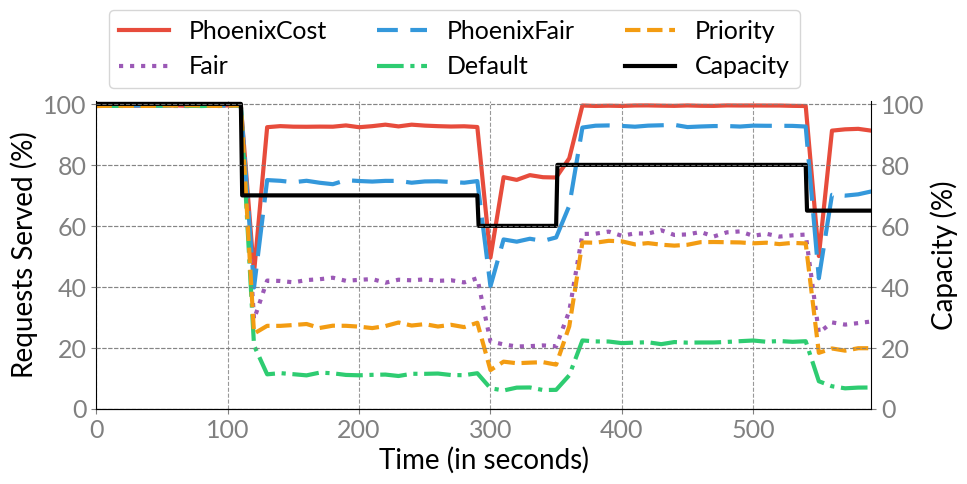} &
    \includegraphics[width=0.3\textwidth]{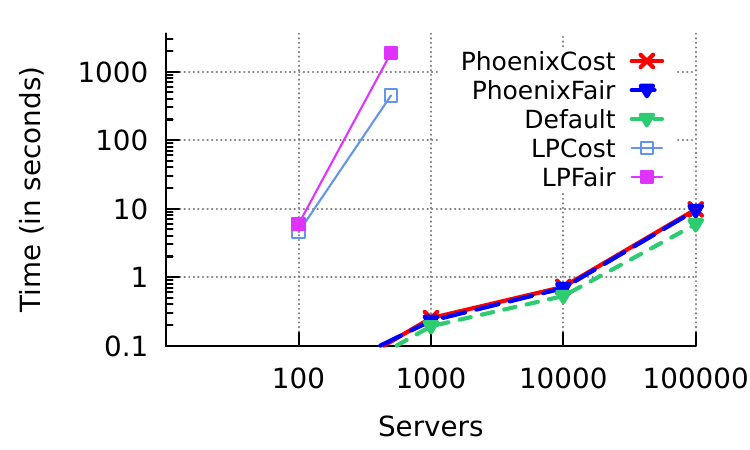} &
    \includegraphics[width=0.3\textwidth]{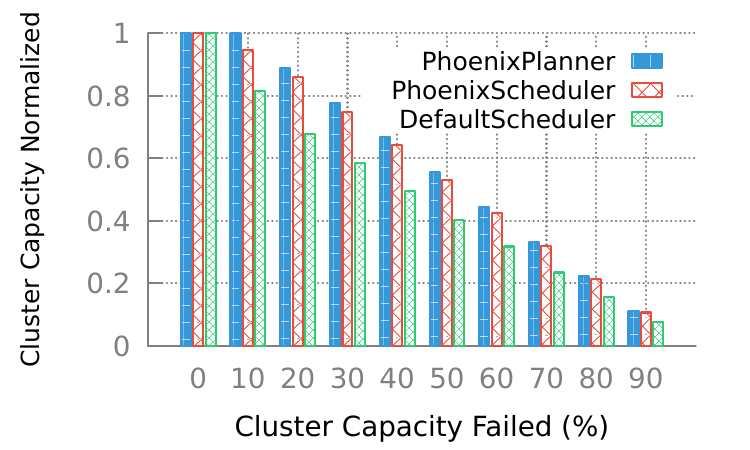} \\
    (a) & (b) & (c) \\
    \end{tabular}
    \caption{\small (a) As the cluster capacity varies significantly over a span of 10 minutes, Phoenix can recover quickly and serve nearly 2$\times$ user requests compared to baselines. Simulation on a 10,000-node cluster by replaying real-world Alibaba traces. (b) AdaptLab benchmarking on a Linux machine with 24 cores. Phoenix is only slightly slower than Default. The LP does not scale beyond 1000 nodes. (c) \textbf{Breakdown of Phoenix performance.} Loss of utilization with only Phoenix planner compared with both planner and scheduler enabled shows that both modules are highly efficient.}
    \label{fig:misc}
\end{figure*}

We evaluate Phoenix at scale using the AdaptLab simulator and report the observations, with all results averaged across 5 trials. Here, we report findings from the P90 service-level criticality tagging scheme and the CPM-based resource allocation model. Additional results are available in Appendix~\ref{appendix:eval}. The baselines, LPCost and LPFair, are excluded due to poor scalability (refer Figure~\ref{fig:misc}(b)).

\parab{Cooperative degradation improves the overall critical service availability.} In Figure \ref{fig:freq_p90}(a), we report critical service availability ($C_1$ containers activated) across applications under different failure rates. We normalize the availability with respect to the unaffected cluster state and report the average across all applications at each failure level. We observe that Phoenix's cooperative degradation outperforms the two non-cooperative degradation strategies, Fair and Priority. Priority performs poorly because it lacks an operator-level signal of inter-app prioritization, which results in a few applications with many high-criticality microservices using most of the resources. Fair's fairness-aware resource allocation leads to better availability than Priority, yet it suffers performance deterioration due to a lack of criticality awareness. Phoenix, with intra- and inter-app prioritization, offers high availability. Default (which lacks criticality, dependency, or packing efficiency awareness) performs the worst.

\parab{Phoenix maximizes operator-level objectives.} Figure~\ref{fig:freq_p90}(b) reports normalized revenue with respect to the state before failure. PhoenixCost offers superior performance due to its efficient packing while explicitly maximizing revenue. The fairness-based schemes perform poorly. Figure~\ref{fig:freq_p90}(c) shows the deviations from fair share across three failure levels of 10\%, 50\%, and 90\%. Ideally, the deviation from fair share must be zero. A negative fair share occurs when an application receives fewer resources than its max-min fair allocation, and a positive fair share when it receives more. With varying failure rates, PhoenixFair has the lowest total deviation. Due to the indivisibility of microservices within an application and the inability to activate beyond fair-share, Fair has a high negative deviation. Since PhoenixFair follows a relaxed fair share criterion, it can achieve lower deviation on both sides. Other schemes perform poorly due to the inability to enforce inter-app fairness.

\parab{Application can meet their critical RTOs under cloud failures with Phoenix's Cooperative Degradation}: Figure~\ref{fig:misc} (a) reports the requests served (y-axis) vs. time (x-axis) by replaying Alibaba traces in AdaptLab. As capacity varies (shown by the solid black line), we see the benefits of Phoenix's cooperative degradation over its non-cooperative counterparts. Phoenix serves $2\times$ requests in comparison to non-cooperative baselines: Fair and Priority.

\parab{Phoenix scales well to real-world cluster sizes.} Figure~\ref{fig:misc} (b) reports the time overheads incurred by Phoenix and baselines on a Linux machine with 24 physical cores and 48 logical processors. LP variants do not scale beyond 1000-server clusters, even with applications with less than 20 microservices. Phoenix's time overheads are comparable to Default, taking less than 10 seconds on 100,000 servers while handing large application sizes up to 3000 microservices.

\parab{Phoenix is resource-efficient.} Figure \ref{fig:misc} (c) shows the cluster capacity utilized under various failure rates by Phoenix planner, Phoenix scheduler (output obtained after planner and scheduler), and the Default scheduler. We observe that the Phoenix scheduler consistently outperforms the default scheduling of Kubernetes, demonstrating superior packing efficiency. Moreover, the drop in cluster utilization from planner to scheduler output is minimal.

\noindent
\vspace{4mm}
In summary, we make the following observations:
\vspace{-2mm}
\begin{itemize}[leftmargin=*,nolistsep]
    \item Phoenix's cooperative degradation approach provides high availability and targeted recovery of critical services across applications while maximizing operator objectives compared to non-cooperative approaches (Figures ~\ref{fig:scatter} and ~\ref{fig:freq_p90})
    \item Phoenix's design of diagonal scaling is practical with today's applications, using Overleaf as an example. (Fig.~\ref{fig:phoenix-cloudlab})
    \item Phoenix scales well to real-world data center sizes. (Fig.~\ref{fig:freq_p90})

\end{itemize}

\section{Discussion and Limitations}
\label{sec:practical}

Phoenix takes the first steps towards application-agnostic automated resilience management in public clouds. However, several challenges remain to be solved in this setting.

\parab{Stateful Workloads}: Phoenix is currently restricted to stateless workloads, based on the premise that containers can be safely terminated and restarted to resume serving requests. Expanding these degradation controls to include stateful workloads poses substantial challenges, primarily the need to reliably persist state across container restarts. 

\parab{Large-scale adoption}: To achieve broad adoption, Phoenix needs to be complemented with a modular application design that can be easily decoupled at deployment time into critical and non-critical workloads. Future work should focus on (1) applying learning-based methods to system logs to modularize existing applications by distinguishing critical from non-critical components, and (2) creating container runtime frameworks that delegate packaging and deployment decisions to automated platforms,  such as Service Weaver~\cite{ghemawat2023towards}.

\parab{Other degradation modes}: Phoenix's container-level degradation is orthogonal to other degradation modes supported by various applications, such as request-level shedding and Quality of Service (QoS) degradation. Phoenix can be combined with these complementary resilience solutions in the future to achieve better overall efficiency.

\parab{Dynamic Criticality Tagging}: Phoenix currently employs static criticality tags for containers. Some applications may benefit from dynamic tags that adjust based on contextual factors, such as time of day or user behavior, to guide real-time decisions on microservice degradation. Future work could expand Phoenix by introducing criticality tagging APIs that allow applications to assign criticality tags dynamically.

\parab{Dynamic Resource Profiling}: Phoenix relies on deployment specifications (such as YAML or TOML) to estimate the capacity freed during degradation. However, degrading user-facing services can influence user behavior, which in turn can change resource demands. This presents an opportunity to extend Phoenix's design with a learning mechanism to adapt to changing resource profiles~\cite{bhardwaj2023cilantro, qiu2020firm, rzadca2020autopilot}.

\parab{Adversarial or Incorrect Criticality Tags}: In dynamic environments with frequent container additions and updates, developers may employ pre-deployment checks using chaos tests to verify tagging and prevent the deployment of incorrect criticality tags. Furthermore, independent tools that can verify the correctness of criticality tags at the application level and the robustness of operator objectives at the infrastructure level can be devised in the future. Operators can employ policies such as resource fairness to limit the impact of incorrect tags.

\vspace{-1mm}
\section{Conclusion}

In this paper, we introduce, diagonal scaling, a cooperative graceful degradation technique that involves turning off non-critical containers to mitigate the impact of large-scale failures in public clouds. We design an automated resilience management system, Phoenix, that leverages diagonal scaling with criticality tags to simultaneously meet application resilience requirements and operator objectives.  We lay the foundation for future research in cloud resilience management with our open-source platform. We identify several open challenges in this space: developing automated resilience management in stateful settings, dynamic criticality tagging, learning-based resource saving estimation, etc.

\section{Acknowledgements}

We sincerely thank our shepherd, Tianyin Xu, and the anonymous reviewers for their
insightful feedback. Additionally, we thank Sujata Banerjee, Lalith Suresh, Saurabh Bagchi, and Sharad Agarwal for their early feedback, which helped shape our paper.

\bibliographystyle{unsrt}
\balance
\bibliography{sample}
\clearpage

\appendix

\appendix
\section{Artifact Appendix}

\subsection{Abstract}

We make our code available on GitHub, including the automated resilience management system Phoenix and the resilience benchmarking platform AdaptLab. 

Our artifact enables users to deploy Phoenix on a Kubernetes cluster with two applications: Overleaf, a real-world collaborative document editing application, and HotelReservation from DeathStarBench. We include additional scripts for a Cloudlab-specific deployment. 

AdaptLab, our benchmarking platform, can emulate realistic public cloud environments of diverse cluster sizes. AdaptLab is a comprehensive testbed for benchmarking various resilience solutions under different failure rates, offering support for key metrics such as critical service availability, operator metrics such as cluster utilization, fairness, and revenue, and systems overheads such as time for adaptation. Our AdaptLab demonstration uses real-world microservice dependency graphs derived from Alibaba cluster traces.  Finally, AdaptLab is extensible and can support the development and testing of new degradation policies, providing a foundation to advance cloud resilience research.

\subsection{Artifact check-list (meta-information)}

{\small
\begin{itemize}
  \item {\bf Program:} Phoenix, our automated resilience management system, and AdaptLab, our resilience benchmarking platform.
  \item {\bf Compilation:} Compatible with Python 3.10 (or higher), and tested with Python 3.10.
  \item {\bf Data set:} Alibaba cluster microservices traces 2021.
  \item {\bf Run-time environment:} Phoenix deployed atop a Kubernetes cluster on CloudLab. AdaptLab is tested on a Linux machine running Ubuntu 20.04.4 LTS with 48 cores.
  
  \item {\bf Hardware:} AdaptLab experiments are conducted on a Linux machine with 48 cores running Ubuntu 20.04.4 LTS. Overleaf and HotelReservation experiments are conducted on a CloudLab cluster of 25 nodes with d710 machines.
  
  \item {\bf Metrics:}  We test Phoenix and baselines on two operator-level objectives: 1) Revenue and 2) Fairness. We introduce critical service availability to measure resiliency scores for each application. We also measure system overheads such as time taken to determine the target cluster state, packing efficiency, etc.
  \item {\bf Experiments: } 
  Scripts are available in the \texttt{plotscripts} folder. Detailed instructions are in the \texttt{README.md} file. 
  \item {\bf How much disk space is required (approximately)?: } 15 GB.
  \item {\bf How much time is needed to prepare workflow (approximately)?: } 30 minutes.
  \item {\bf How much time is needed to complete experiments (approximately)?: } 5-7 hours.
  \item {\bf Publicly available?: } Yes.
  \item {\bf Code licenses (if publicly available)?: } Yes, Apache License version 2.0.
  \item {\bf Archived (provide DOI)?:} Yes, we have archived our code on Zenodo: \url{https://doi.org/10.5281/zenodo.14483674}

\end{itemize}
}

\subsection{Description}

\parab{How to access}: Phoenix's source code is available on GitHub: \url{https://github.com/NetSAIL-UCI/Phoenix}

\parab{Hardware dependencies:} While there are no special hardware requirements for running AdaptLab, the experiments were conducted on an Intel(R) Xeon(R) Gold 6246 CPU @ 3.30GHz Linux machine running Ubuntu 20.04.4 LTS with 48 cores. Next, we require a CloudLab cluster of 25 d710 machines to test the Phoenix Controller on real-world microservices.  

\parab{Software dependencies:} Gurobi Optimizer and Apache Spark are required. The installation instructions are specified in the GitHub repository.

\parab{Datasets:} We analyze the Alibaba cluster microservice traces (2021) and extract 18 applications following the methodology cited in Luo et al., 2022 [87].
The dataset can be found here: \url{https://github.com/alibaba/clusterdata/tree/master/cluster-trace-microservices-v2021}

\subsection{Installation}

Please clone the GitHub repository and install Phoenix's dependencies. We have prepared the required Python packages. Please run the script to install the packages as follows:
\texttt{pip install requirements.txt}

Users should also install the Gurobi optimizer and obtain the Gurobi license following instructions in this link: \url{https://www.gurobi.com/features/academic-named-user-license/}. 

The Apache Spark package is required for Alibaba trace extraction (however, this step is not critical for reproducing results in this paper). Steps for installing Apache Spark can be found here: \url{https://spark.apache.org/downloads.html}.

\subsection{Experiment workflow}

Phoenix is evaluated in two settings: 1) \textit{Real world}: a 25-node CloudLab Kubernetes cluster running five instances of two microservice applications, and 2) \textit{AdaptLab Simulation}: a  100K-node simulated cluster with Alibaba cluster traces. In the CloudLab setting, we simulate a failure where 58\% of the cluster is failed. We then measure Phoenix's performance on its ability to maintain critical service availability and compare it with various baselines.

Next, we simulate a real-world public cloud cluster with 100,000 nodes running real-world microservice application dependency graphs obtained using Alibaba traces. This experiment demonstrates how, at a large scale, Phoenix is able to outperform other baselines, showing Phoenix's efficacy in real-world cloud environments. 
\subsection{Evaluation and Expected Results}

We conduct a comprehensive evaluation of Phoenix across real-world and simulation settings. First, we evaluate Phoenix's ability to maximally satisfy the application's resilience objective using the Critical Service Availability metric. We expect Phoenix's performance to be superior to the baselines. Second, we assess Phoenix's effectiveness in achieving operator objectives using two objectives: cost and fairness. Phoenix should be able to outperform other baselines. Finally, we measure Phoenix's time overhead and expect it to be within seconds for real-world size clusters (100K servers).

\begin{algorithm}[ht]
\DontPrintSemicolon
\SetKwInput{KwInput}{Input}                %
\SetKwInput{KwOutput}{Output}   
\KwInput{\textbf{\textit{P}}: Planner Outputted List, \textbf{\textit{A}}: Assignments}
\KwOutput{\textbf{\textit{A}}: Final assignment of microservice to server}
\KwData{\textbf{\textit{C}}: Cluster State object}
\SetKwFunction{FRepack}{RepackMicroservicesToFit}
\SetKwFunction{FMain}{Main}
\SetKwFunction{FBestFit}{GetBestFit}
\SetKwFunction{FDelete}{DeleteLowerRanksToFit}
\SetKwFunction{FUpdate}{UpdateClusterState}
\SetKwFunction{FMigrateFeasible}{IsMigrationFeasible}
\SetKwFunction{FMigrate}{RepackMicroservices}
\SetKwFunction{FDeletePod}{DeletePod}
\SetKwProg{Fn}{def}{:}{}

\Fn{\FMain}{
    \ForEach{$ms \in P$}{
        \If{$ms \notin A$}{
            node = \FBestFit($ms$, $A$)\;
            \lIf{$node == \text{None}$}{node = \FRepack($ms$, $A$)}
            \lIf{$node == \text{None}$}{node = \FDelete($ms$, $A$)}
            \lIf{$node == \text{None}$}{\KwRet None}
            \lElse{\FUpdate($node$, $A$)}
        }
    }
    \KwRet $A$ \;
}

\SetKwProg{Fn}{def}{:}{}
\Fn{\FBestFit{$ms$, $A$}}{
    best = \text{None}\;
    \ForEach{$node \in C.\text{nodes}$}{
        \lIf{$node.\text{size} \geq ms.\text{size} \land node.\text{empty} \leq \text{best.empty}$}{
            best = $node$
        }
    }
    \KwRet best \;
}

\SetKwProg{Fn}{def}{:}{}
\Fn{\FRepack{$ms$, $A$}}{
    \ForEach{$node \in \text{sorted}(A.\text{nodes}(\text{key}=\text{RemainingCapacity}, \text{reverse=True}))$}{
        \lIf{\FMigrateFeasible($node$)}{
            \FMigrate($node$)\;
            \KwRet $node$ \;
        }
    }
    \KwRet \text{None} \;
}

\SetKwProg{Fn}{def}{:}{}
\Fn{\FDelete{$ms$, $A$}}{
    \ForEach{$loms \in \text{sorted}(A.\text{podsRank}, \text{reverse=True})$}{
        node = \FDeletePod($loms$)\;
        \lIf{$ms.\text{size} \leq node.\text{size}$}{
            \KwRet $node$
        }
    }
    \KwRet \text{None} \;
}
\caption{Packing Heuristic}
\label{hscheduler}
\end{algorithm}

\section{Packing Heuristic}
\label{sched:code}
We now provide the pseudocode for the bin-packing heuristic in Algorithm \ref{hscheduler}. The main function runs in the following manner: First, it traverses over the list of microservices outputted by the planner. If a microservice is not scheduled, it first calls the best-fit subroutine~\cite{bestfit} to find the node with the least remaining capacity to accommodate the microservice. If a node is found, we proceed to line 8 to update the cluster state and proceed to the next microservice. If a node is not found, we proceed to the repacking strategy. The repacking strategy iterates over the nodes (in the order of most empty to least empty). It also checks the feasibility of migrating smaller-sized microservices onto other target nodes. If such a node is found that can free up capacity by repacking smaller microservices to other nodes then it migrates, as shown in Line -17, and returns the candidate node else it returns None, i.e., it cannot find any node in the list. 

Finally, if repacking also does not work then we perform deletions of lower-ranked microservices to free up capacity for the higher-ranked microservice to be scheduled. If such a node is freed up then we return the node else we return none. 

\section{Linear Programming Formulations}
\label{appendix:lp}

We discuss two operator objectives and their corresponding Integer Linear Programming (ILP) formulations.

\parab{Fairness Objective}: We first discuss a max-min fairness approach---ensuring each application receives an equitable share of resources---the ILP will be as follows:
\vspace{-1mm}
\begin{align*}
    Maximize \; F
\end{align*}
\begin{flalign}
\sum_{j}R_{ij}*x_{ij} > F \;| \;\forall app_{i}, \;\forall m_j \;\epsilon \; app_i
\end{flalign}

However, this objective does not work correctly when resource demands across applications are skewed. For example, if three applications have resource demands of 10, 50, and 90 units and the existing resources are 100 units then the above LP can deem 10, 10, and 80 units as an acceptable configuration. In summary, this objective alone may lead to unfair distribution of the excess resources after satisfying the minimum application.

One approach is to recursively call the above ILP formulation to ensure that excess resources are distributed fairly. However, this approach, which requires several ILP calls, can be extremely compute-intensive.

To make this more efficient, we pre-compute the water-fill fair share outside of the ILP and then pass this water-fill fair share for each app to the ILP as parameters. The resulting mathematical formulation of which is as follows:
\vspace{-1mm}
\begin{align*}
    Maximize \; F
\end{align*}
\begin{flalign}
\sum_{j}R_{ij}*x_{ij} \geq F \;| \;\forall app_{i}, \;\forall m_j \;\epsilon \; app_i\\
\sum_{j}R_{ij}*x_{ij} \leq FS_i \;| \;\forall app_{i}, \;\forall m_j \;\epsilon \; app_i
\end{flalign}

$FS_i$ refers to the water-fill fair share for application $app_i$ which is passed as the parameter to the above ILP. The time complexity introduced by the pre-computation step of water-fill is negligible in comparison to solving the ILP and does not introduce any additional bottlenecks.

\parab{Revenue Based}: The cloud operator's primary objective is to maximize the revenue and therefore, we rely on the willingness-to-pay of each application per unit resource.

\vspace{-1mm}
\begin{align*}
    Maximize \; \sum_{i} \sum_{j} C_i*x_{ij}*R_{ij} 
\end{align*}

where $C_i$ represents the revenue the cloud generates by turning on 1 unit of resource of application $app_i$.

\section{Extending to Multiple Replicas}
\label{appendix:replicas}

A large number of applications run multiple containers in a microservice deployment. In the paper, we discuss Phoenix where each microservice has a single container. Here, we provide a mechanism to extend it such that each microservice can have multiple replicas, as is common practice. Towards this, the Phoenix Planner algorithm requires no changes as it does not operate on the microservice instance information but only requires the resource information as a whole. On the bin-packing heuristic side, we can easily modify our implementation to treat a microservice as active or not if all microservices replicas are activated. Specifically, when we obtain a list from the planner, we traverse based on the ranking of microservices, for every microservice we see whether all the replicas are running or not. If not we assign them using the three-pronged bin-packing heuristic strategy of best-fit, repacking, and deletion. If one of the replicas does not fit, we do not extend beyond and terminate the for loop by deleting all microservice replicas. 

\section{Microservice Deletion, Migration, and Restarts in Kubernetes}

In this subsection, we detail the fine-grained processes involved in microservice deletion, migration, and restarts at the Kubernetes level when the Phoenix agent issues the corresponding commands.

\parait{Deletion}: When a graceful shutdown request is issued, the cluster scheduler starts by removing service endpoints on each server where the microservice is running to avoid any incoming requests. Next, the cluster scheduler scales the replicas gradually down by issuing SIGTERM commands. These containers then complete any outstanding requests and then terminate. However, if a container doesn't terminate after a threshold time limit is reached, a SIGKILL command request is issued.  Currently, Phoenix Agent supports draining two types of traffic: 1. Stateless Connections and 2. Websocket connections. In addition to the deletion of microservices, Phoenix Agent is extensible for running other user-defined subroutines pre-deletion or post-deletion to manage upstream or downstream dependencies. These can include toggling knobs or feature flags \cite{meza2023defcon, featureflags} in the upstream or downstream services to dynamically adjust the behavior of upstream/downstream microservice in run-time without rebuilding binaries. It is important to note that the Phoenix Agent only supports executing operations on the cluster level and does not perform any application-level changes.

\parait{Migrations}: Migrations typically involve the replacement of a microservice from one server to another to improve the resource efficiency in the cluster. However, depending on cluster operators, we make migrations optional because migrations can incur additional latency overheads, increasing the time to reach the desired state. Migrations in Phoenix are currently performed in two stages currently, first by restarting the microservice on the target server, reconfiguring the service, and then deleting the old microservice. We employ a similar traffic-draining strategy by rerouting the requests to the new instantiated microservice. 

\parait{Restarts}: Restarting a microservice means that the microservice was impacted as a result of a failure event. We perform restarting in a few steps. We begin by cleaning up stale entries on the ip tables. We then instantiate the container(s) on the desired servers first. Once the containers are running, we reconfigure the service to start serving traffic. Similar to deletion, the Phoenix agent is extensible to perform any pre-restart or post-restart subroutines that application developers provide.

\section{Detailed Results}
\label{appendix:eval}

\subsection{Cloudlab Resource Distribution}
\label{appendix:cloudlab-resource-dist}

 \begin{figure}
    \centering
    \includegraphics[width=0.33\textwidth]{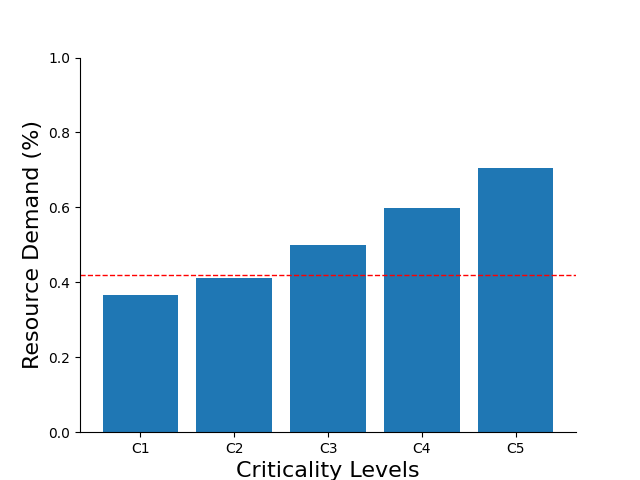}
    \caption{\small \textbf{Breakdown of resources across criticalities for the real-world experiment.}}
    \label{fig:r-vs-c}
\end{figure}

We report the aggregate resource consumption per criticality level across 5 instances (Overleaf0, Overleaf1, Overleaf2, HR0, HR1) on a CloudLab cluster with 200 CPUs. The resource division between $C_1$ and non-critical ($C_2$ and lower) is  $\approx$ 60:40 as shown in figure~\ref{fig:r-vs-c}. Furthermore, the load is such that all $C_1$ require $\approx$40\% of cluster capacity, and all applications together require $\approx$70\% of cluster capacity. For all cloudlab experiments reported in \S\ref{sec:eval}, we introduce failure rates up to $\approx$42\%, below which $C_1$ microservices could fail---the highest acceptable degradation. 

\subsection{Standalone Testing}
\begin{figure*}[h]
    \centering
    \begin{tabular}{ccc}
    \includegraphics[width=0.31\textwidth]{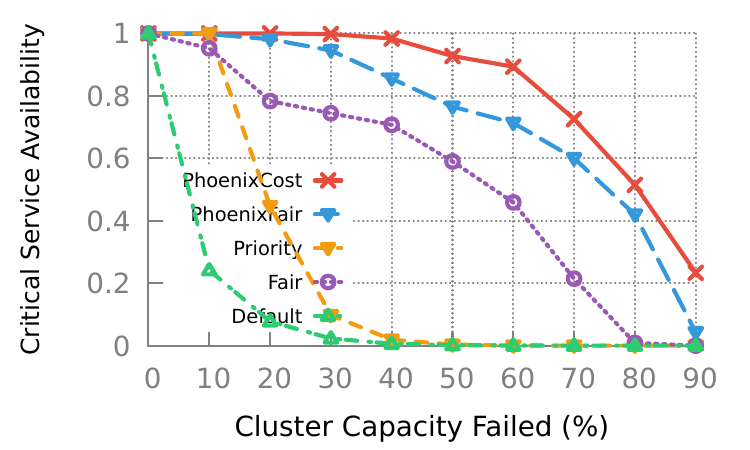} &
    \includegraphics[width=0.31\textwidth]{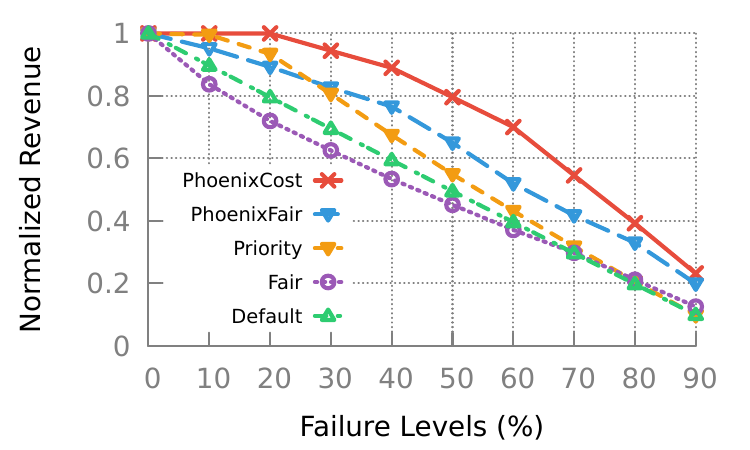} &
    \includegraphics[width=5cm,height=3.2cm]{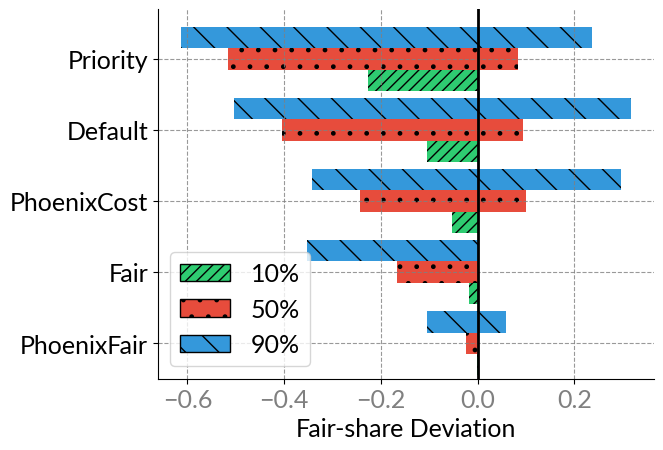} \\
    \end{tabular}
    \caption{\small \textbf{Resilience schemes evaluated on Alibaba with Service-Level-P50 criticality tagging scheme and CPM-based resource assignment scheme on a 100000-node cluster} (a) Aggregate critical service availability across an application at different capacity failure scenarios shows that PhoenixFair and PhoenixCost activate more paths consistently than baselines. (b) Revenue achieved. (c) Fair-share deviation with respect to water-fill fairness. }
    \label{fig:svcp50}
\end{figure*}

\begin{figure*}[h]
    \centering
    \begin{tabular}{ccc}
    \includegraphics[width=0.31\textwidth]{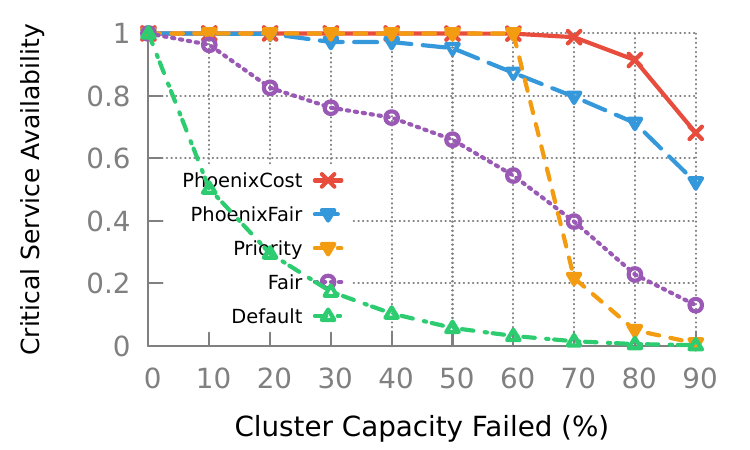} &
    \includegraphics[width=0.31\textwidth]{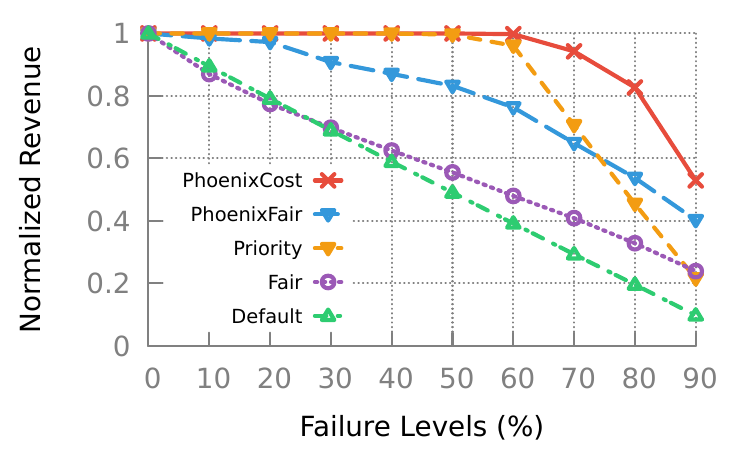} &
    \includegraphics[width=5cm,height=3.2cm]{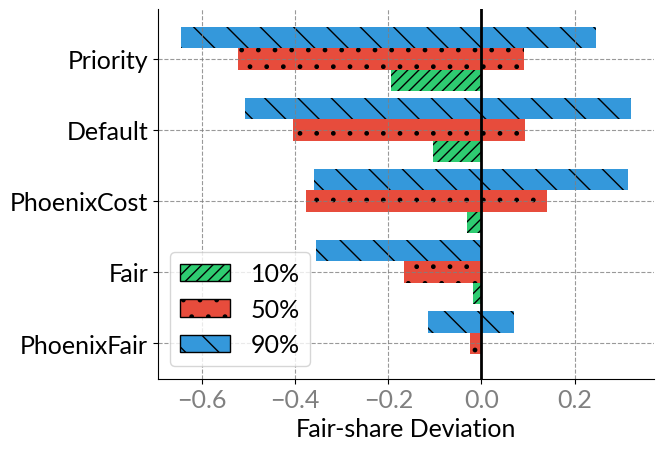} \\
    \end{tabular}
    \caption{\small \textbf{Resilience schemes evaluated on Alibaba with Frequency-Based-P50 criticality tagging scheme and CPM based resource assignment scheme on a 100000-node cluster} (a) Aggregate critical service availability across an application at different capacity failure scenarios shows that PhoenixFair and PhoenixCost activate more paths consistently than baselines. (b) Revenue achieved. (c) Fair-share deviation with respect to water-fill fairness. }
    \label{fig:freqp50}
\end{figure*}

\begin{figure*}[h]
    \centering
    \begin{tabular}{ccc}
    \includegraphics[width=0.31\textwidth]{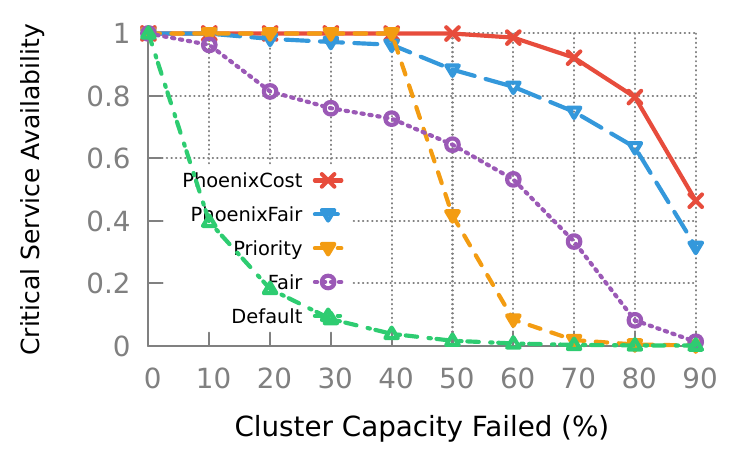} &
    \includegraphics[width=0.31\textwidth]{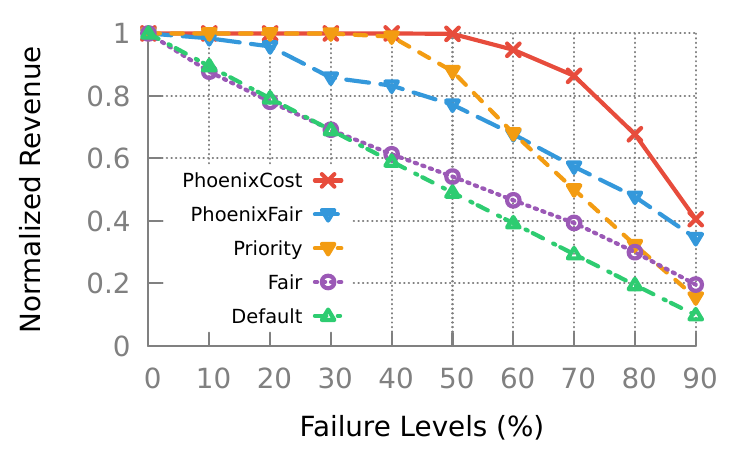} &
    \includegraphics[width=5cm,height=3.2cm]{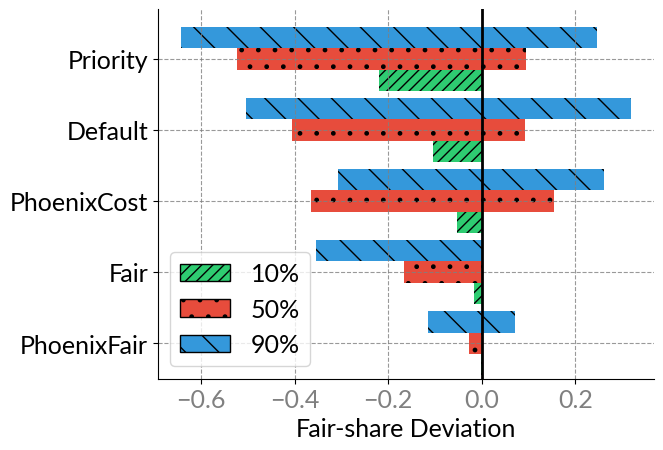} \\
    \end{tabular}
    \caption{\small \textbf{Resilience schemes evaluated on Alibaba with Frequency-Based-P90 criticality tagging scheme and CPM based resource assignment scheme on a 100000-node cluster} (a) Aggregate critical service availability across an application at different capacity failure scenarios shows that PhoenixFair and PhoenixCost activate more paths consistently than baselines. (b) Revenue achieved. (c) Fair-share deviation with respect to water-fill fairness.}
    \label{fig:freqp90}
\end{figure*}

\begin{figure*}[h]
    \centering
    \begin{tabular}{ccc}
    \includegraphics[width=0.31\textwidth]{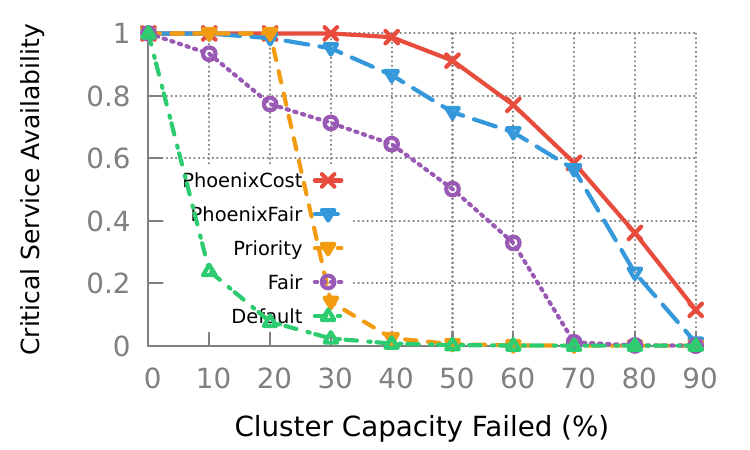} &
    \includegraphics[width=0.31\textwidth]{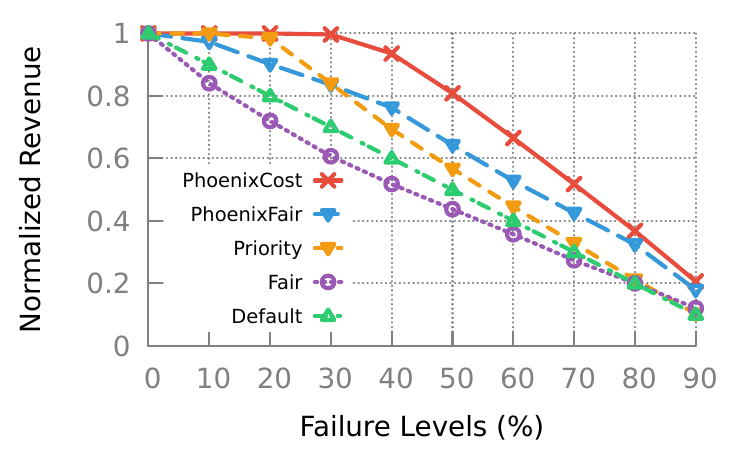} &
    \includegraphics[width=5cm,height=3.2cm]{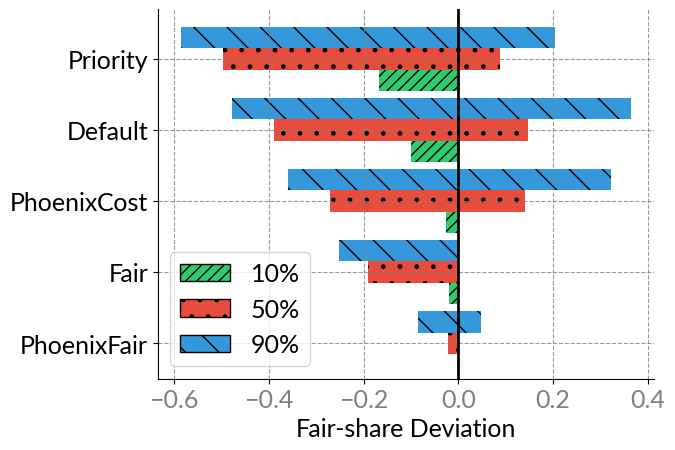} \\
    \end{tabular}
    \caption{\small \textbf{Resilience schemes evaluated on Alibaba with Service-Level-P50 criticality tagging scheme and LongTailed based resource assignment scheme on a 100000-node cluster} (a) Aggregate critical service availability across an application at different capacity failure scenarios shows that PhoenixFair and PhoenixCost activate more paths consistently than baselines. (b) Revenue achieved. (c) Fair-share deviation with respect to water-fill fairness. }
    \label{fig:svcp50longtailed}
\end{figure*}

\begin{figure*}[h]
    \centering
    \begin{tabular}{ccc}
    \includegraphics[width=0.31\textwidth]{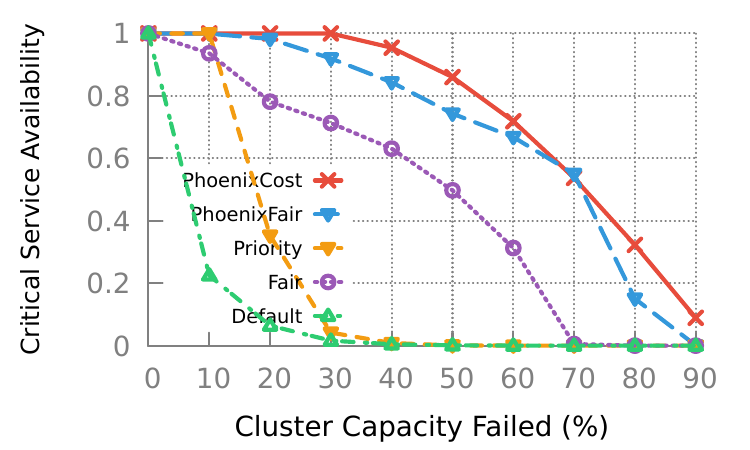} &
    \includegraphics[width=0.31\textwidth]{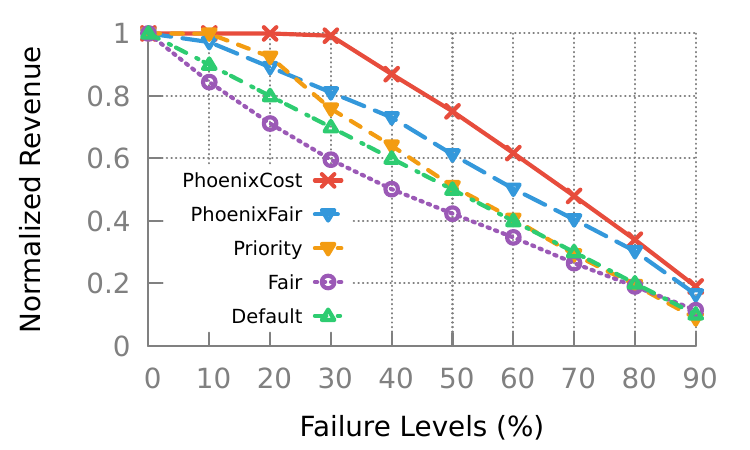} &
    \includegraphics[width=5cm,height=3.2cm]{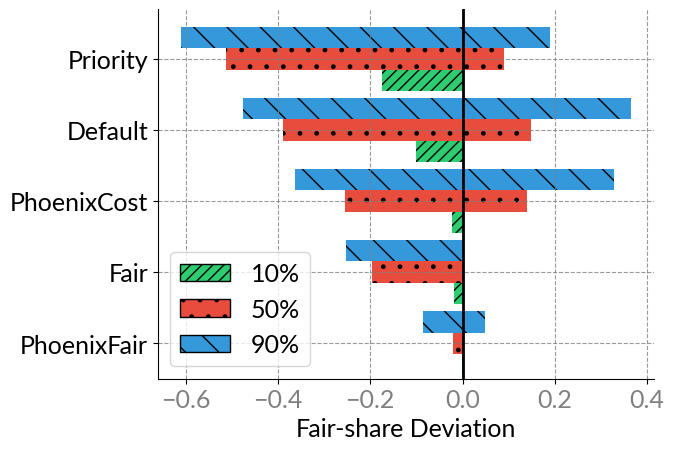} \\
    \end{tabular}
    \caption{\small \textbf{Resilience schemes evaluated on Alibaba with Service-Level-P90 criticality tagging scheme and LongTailed based resource assignment scheme on a 100000-node cluster} (a) Aggregate critical service availability across an application at different capacity failure scenarios shows that PhoenixFair and PhoenixCost activate more paths consistently than baselines. (b) Revenue achieved. (c) Fair-share deviation with respect to water-fill fairness. }
    \label{fig:svcp90longtailed}
\end{figure*}

\begin{figure*}[h]
    \centering
    \begin{tabular}{ccc}
    \includegraphics[width=0.31\textwidth]{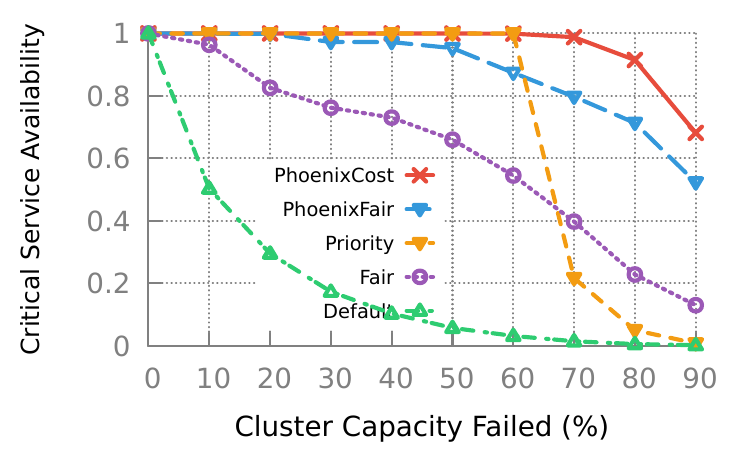} &
    \includegraphics[width=0.31\textwidth]{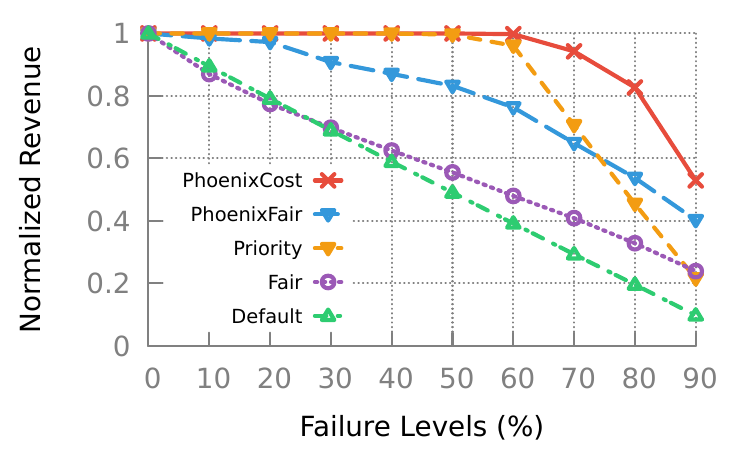} &
    \includegraphics[width=5cm,height=3.2cm]{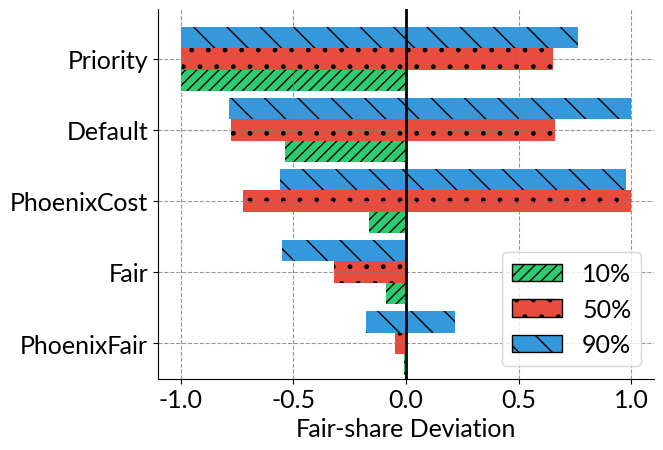} \\
    \end{tabular}
    \caption{\small \textbf{Resilience schemes evaluated on Alibaba with Freq-Based-P50 criticality tagging scheme and LongTailed based resource assignment scheme on a 100000-node cluster} (a) Aggregate critical service availability across an application at different capacity failure scenarios shows that PhoenixFair and PhoenixCost activate more paths consistently than baselines. (b) Revenue achieved. (c) Fair-share deviation with respect to water-fill fairness. }
    \label{fig:freqp50longtailed}
\end{figure*}

\begin{figure*}[h]
    \centering
    \begin{tabular}{ccc}
    \includegraphics[width=0.31\textwidth]{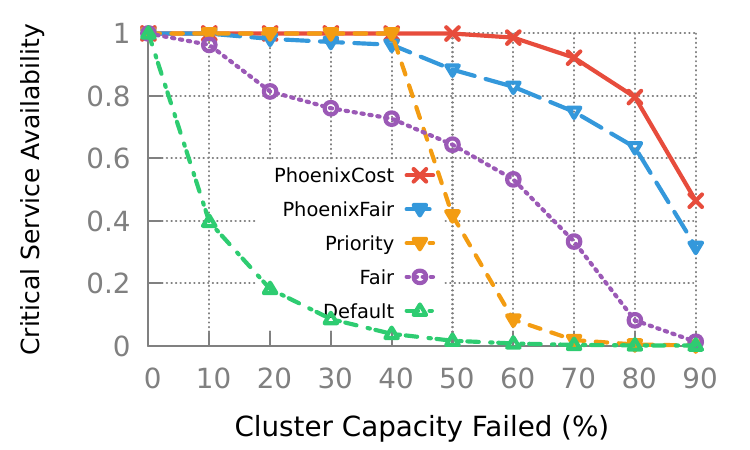} &
    \includegraphics[width=0.31\textwidth]{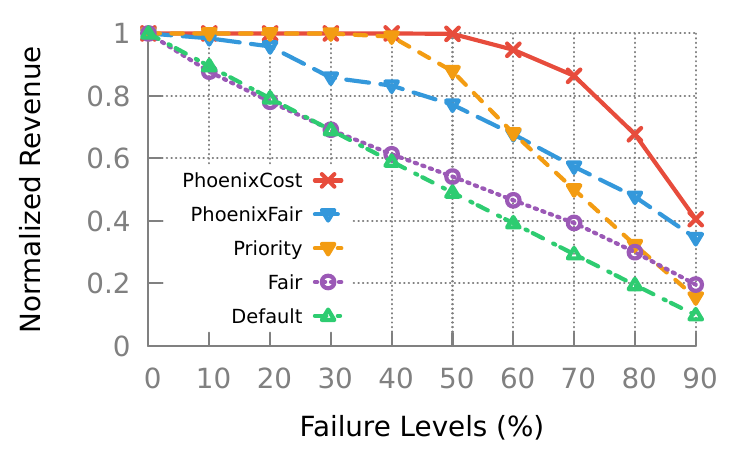} &
    \includegraphics[width=5cm,height=3.2cm]{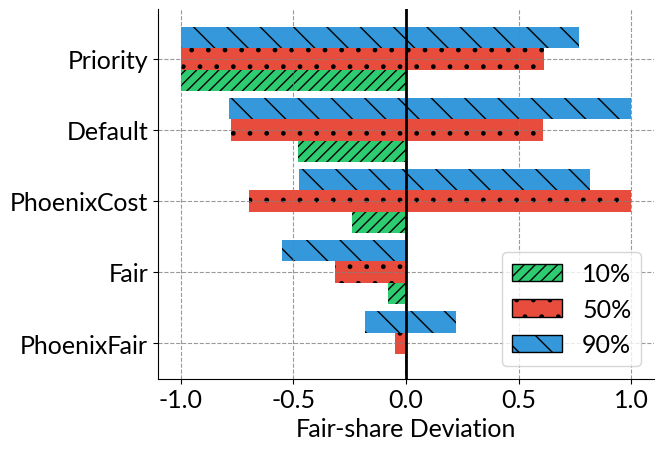} \\
    \end{tabular}
    \caption{\small \textbf{Resilience schemes evaluated on Alibaba with Freq-Based-P90 criticality tagging scheme and LongTailed based resource assignment scheme on a 100000-node cluster} (a) Aggregate critical service availability across an application at different capacity failure scenarios shows that PhoenixFair and PhoenixCost activate more paths consistently than baselines. (b) Revenue achieved. (c) Fair-share deviation with respect to water-fill fairness. }
    \label{fig:freqp90longtailed}
\end{figure*}

In this section, we list down the results for the two resource models we considered and four criticality tagging schemes when running Alibaba DGs on a 100,000-node server on the simulator. In summary, we get 4 (criticality tagging schemes) * 2 resource schemes that we report here. Consistently, in all cases Figures~\ref{fig:svcp50},~\ref{fig:freqp50},~\ref{fig:freqp90},~\ref{fig:svcp50longtailed},~\ref{fig:svcp90longtailed},~\ref{fig:freqp50longtailed},~\ref{fig:freqp90longtailed}, Phoenix outperforms other baselines.

\label{appendix:synthetic}

\section{Alibaba Analysis}
\label{appendix:alibaba}

\begin{figure*}[!ht]
\centering
\begin{subfigure}{.33\textwidth}
    \centering
    \includegraphics[width=0.99\textwidth]{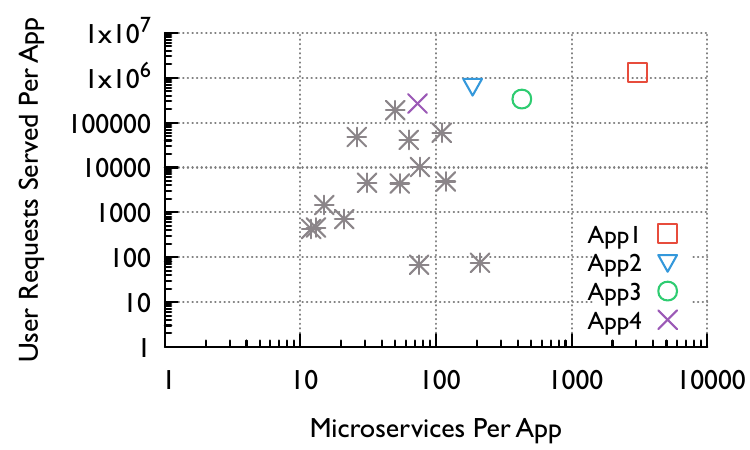}
    \caption{\small Application size vs. user requests served}
    \label{fig:alibaba_ms_scatter}
\end{subfigure}
\begin{subfigure}{.33\textwidth}
    \centering
    \includegraphics[width=0.99\textwidth]{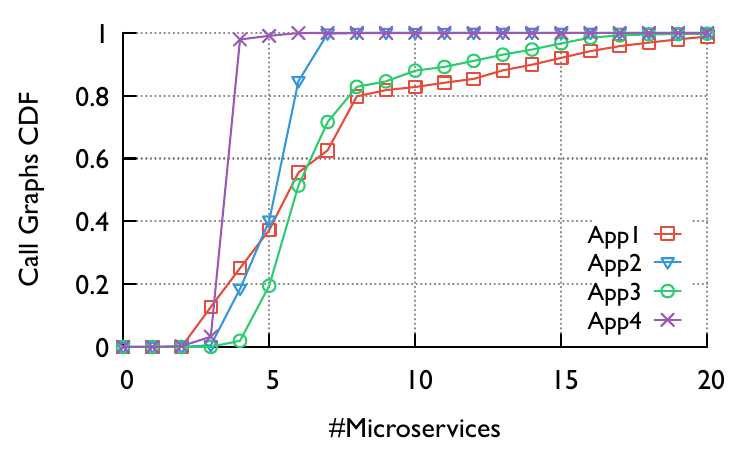}
    \caption{\small Call graph size distribution}
    \label{fig:alibaba_ms_cdf}
\end{subfigure}
    \centering
\begin{subfigure}{.33\textwidth}
    \includegraphics[width=0.99\textwidth]{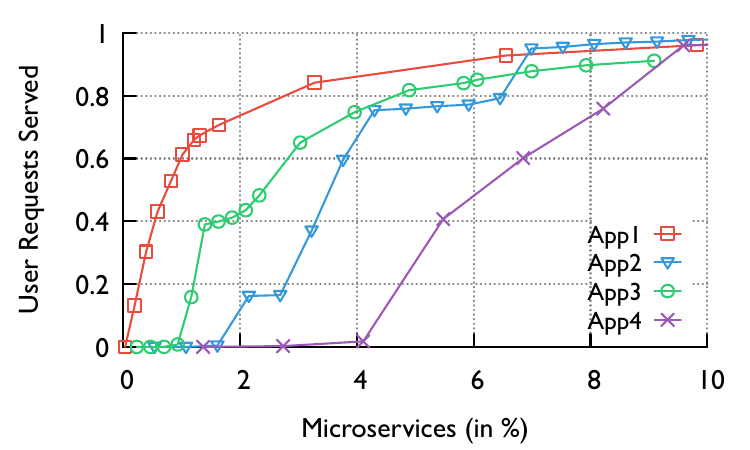}
    \caption{\small Requests served with pruned graph}
    \label{fig:alibaba_CG_users}
\end{subfigure}
\caption{\small \textbf{Analysis of Alibaba workloads}: We analyze a real-world Alibaba dataset~\cite{luo2021characterizing} over a period of 7 days. (a) The distribution of dependency graph size vs. the number of user requests served shows that a large number of applications have small graphs. (b) The call graph size distribution of the top four applications shows that most call graphs are small. (c) A large fraction of user requests can be served by enabling a small fraction of microservices for all applications.}
\label{fig:alibaba}
\end{figure*}

\subsection{Real-World Dependency Graph Analysis}
To better understand real-world application usage patterns, we analyze the microservice dependency graphs dataset from Alibaba~\cite{luo2021characterizing}. The dependencies are characterized using call graphs, where each call graph corresponds to a single user request and is denoted as a directed graph containing all the calls between microservices triggered by that request.  The dataset consists of over twenty million call graphs collected over 7 days from Kubernetes clusters at Alibaba. We analyze data from this dataset and derive the corresponding application dependency graphs~\cite{luo2022depth,detti2023mubench}. Characteristics of the 18 application dependency graphs are shown in Figure~\ref{fig:alibaba}.

In Figure~\ref{fig:alibaba_ms_scatter}, we show that a large number of applications have small dependency graphs with dozens of microservices, and they serve a large fraction of the user requests. In Figure~\ref{fig:alibaba_ms_cdf}, we analyze the call graph size distribution of the top four applications serving the most user requests. We observe that most call graphs (subgraphs of the corresponding application dependency graphs) only contain a small fraction of the microservices in the application DG. For example, in App1 which contains more than $3000$ microservices, over $80\%$ of call graphs contain fewer than $10$ microservices only. However, different call graphs may include different subsets of microservices. Hence, we conduct another analysis to understand the overlap of microservices \textit{across} call graphs for each application.

Using a Linear Program, we estimate the number of call graphs that can be fully activated with a given fixed number of microservices. For each application, we vary the number of microservices that are allowed to be activated and evaluate the maximum fraction of user requests that can be supported at each size. In Figure~\ref{fig:alibaba_CG_users}, we show that most applications can serve a large fraction of users with a small fraction of microservices activated. In App1 which serves over $1,300,000$ requests and contains over $3000$ microservices, more than $80\%$ user requests can be served by enabling only $3\%$ microservices ($90$ microservices in the DG). Similar pattern also holds for smaller applications. For example, App2 with about $50$ microservices can serve $90\%$ of its user requests with less than 10 microservices enabled.  

The long-tailed distribution of application call graphs implies that most applications can continue to serve a large fraction of user requests even when several microservices in their dependency graphs are turned off. We argue that this property of call graphs can be leveraged to improve application availability during capacity crunch scenarios by turning off non-critical microservices.

\section{Impact on Latency}
\label{appendix:latency-impact}

\begin{table}
    \centering
    \small
    \begin{tabular}{c c c c }
     & & \multicolumn{2}{c}{\textbf{P95 Latencies (in ms)}}\\
        \textbf{Applications}& \textbf{Services}&\textbf{Before}& \textbf{After}\\
        \hline
        \multirow{3}{*}{Overleaf} & edits & 141 & 144\\
        & compile & 4317.9 & --\\
        & spell\_check & 2296.7 & -- \\
        \hline
        \multirow{4}{*}{HR} & reserve & 55.33 & 50.11 \\
        & recommend & 47.43 & -- \\
        & search & 53.26 & --\\
        & login & 41.8 & --\\
        \hline
    \end{tabular}
    \caption{\small End-to-end $95^{th}$ percentile (P95) latencies for services in HotelReservation and Overleaf before and after diagonal scaling.}
    \label{tab:p95-latencies}
\end{table}

\parab{Phoenix offers high performance for critical services.} Table \ref{tab:p95-latencies} reports the P95 latencies (in ms) before and after diagonal scaling. The services that were pruned through diagonal scaling are denoted as "--". Overleaf does not exhibit partial pruning because it does not have optional microservices. While the service `reserve' of HR is partially pruned (turning off one downstream microservice), the performance overheads are minimal. This is because HR uses gRPC for inter-process communication, which builds on top of HTTP/2 protocol and leverages its ability to detect failed connections, thereby failing fast without incurring overheads~\cite{grpc-fault-tolerance}.

\end{document}